\documentclass[aps,onecolumn,showpacs,notitlepage]{revtex4-1}
\usepackage{amssymb,lipsum}
\usepackage{graphicx,amsmath}

\newcommand{\bea}{\begin{eqnarray}}
\newcommand{\eea}{\end{eqnarray}}
\newcommand{\beq}{\begin{equation}}
\newcommand{\eeq}{\end{equation}}
\newcommand{\benu}{\begin{enumerate}}
\newcommand{\enu}{\end{enumerate}}

\newcommand{\al}{\alpha}
\newcommand{\be}{\beta}
\newcommand{\ga}{\gamma}
\newcommand{\om}{\omega}
\newcommand{\Om}{\Omega}
\newcommand{\ep}{\epsilon}

\newcommand{\si}{\sigma}

\newcommand{\ham}{\mathcal{H}}

\newcommand{\ptl}{\partial}

\newcommand{\bk}{{\bf k}}
\newcommand{\bq}{{\bf q}}

\newcommand{\br}{{\bf r}}

\begin{document}

\title{
Charge Nematicity and Electronic Raman Scattering in Iron-based Superconductors}

\date{\today}

\author{Y. Gallais}
\email{yann.gallais@univ-paris-diderot.fr}
\affiliation{
Laboratoire Mat\'{e}riaux et Ph\'{e}nom\`{e}nes Quantiques, Universit\'{e} Paris Diderot-Paris 7 \& CNRS,
UMR 7162, 75205 Paris, France
}

\author{I. Paul}
\email{indranil.paul@univ-paris-diderot.fr}
\affiliation{
Laboratoire Mat\'{e}riaux et Ph\'{e}nom\`{e}nes Quantiques, Universit\'{e} Paris Diderot-Paris 7 \& CNRS,
UMR 7162, 75205 Paris, France
}

\begin{abstract}
We review the recent developments in electronic Raman scattering measurements of charge nematic fluctuations in iron-based superconductors.
A simple theoretical framework of a $d$-wave Pomeranchuk transition is proposed in order to capture the salient features of the spectra.
We discuss the available Raman data in the normal state of 122 iron-based systems,
particularly Co doped BaFe$_2$As$_2$, and we show that the low energy quasi-elastic peak, the extracted nematic susceptibility and the
scattering rates are consistent with an electronic driven structural phase transition.
In the superconducting state with a full gap the quasi-elastic peak transforms into a finite frequency nematic resonance, evidences for
which are particularly strong in the electron doped systems. A crucial feature of the analysis is the fact that the electronic Raman signal
is unaffected by the acoustic phonons. This makes Raman spectroscopy a unique probe of electronic nematicity.
\end{abstract}

\maketitle

\newpage
\tableofcontents
\newpage

\section{Introduction.}
\label{sec:intro}

The study of correlated phases of matter obtained a big boost with the discovery of the iron-based
superconductors (Fe SC) in 2008~\cite{Kamihara2008}. These systems are interesting not just because they exhibit superconductivity at
temperatures as high as 55 K, but also because they are a rich playground
where the lattice and the electronic charge, spin and orbital degrees of freedom all play important roles.
This complex interplay invariably leads to competition between various
interesting phases that can be stabilized by varying temperature, by doping carriers and by applying
pressure. Understanding this rich physics is a considerable challenge, and consequently the topic
continues to be an active area of research in condensed matter systems~\cite{Johnston2010,Chubukov-review}.

Unlike the cuprate high temperature superconductors where the parent compounds are Mott insulators, the
Fe SC systems are multi-band and multi-orbital metals at all doping. The undoped and the lightly doped
compounds undergo a structural transition from a high temperature ($T > T_S$) tetragonal unit cell
to a low temperature ($T < T_S$) orthorhombic phase (in the case of Fe$_{1+y}$Te$_{1-x}$Se$_x$ the low-$T$
phase is monoclinic for small $x$) which is followed in close proximity in temperature by a magnetic transition at $T_N$ below which
the system is antiferromagnetic. These transitions are suppressed with electron or hole doping, and beyond a certain doping value the
system becomes superconducting with unusually high transition temperature $T_C$. Initial investigations of the superconductivity and its
origin have focused mainly on the interplay between a stripe-like antiferromagnetism (where the magnetic ordering wave-vector is $(\pi, 0)$
or $(0, \pi)$ in the 1Fe/cell notation) and superconductivity. While the issue is not entirely settled, it is
popularly believed that the fluctuations associated with the stripe antiferromagnetism give
rise to a $s^{\pm}$ superconducting pairing symmetry \cite{Kuroki2008,Mazin2008a,Hirschfeld-review} in most,
but possibly not all \cite{Reid2012}, Fe SC families.

\par
Ever since the reports of strongly anisotropic in-plane transport in the 122 systems \cite{Chu2010} in the
orthorhombic phase, a lot of attention has been given to study the property of nematicity in these materials.
A nematic phase of matter is one which breaks rotational symmetry spontaneously, while preserving translational symmetry. Such phases
have been studied extensively since the seventies in classical soft matter systems~\cite{Prost}, but relatively less is known about
their quantum counterparts in electronic systems. However, their existence has been widely postulated for strongly correlated materials
such as quantum Hall systems, cuprates, bilayer ruthenates \cite{Fradkin-review}, and most recently in the Fe SC. In the presence of a
crystalline lattice, a nematic phase breaks discrete rotational symmetry, and as a consequence the associated order parameter is an
Ising variable.
In the context of the Fe SC, this order parameter is nonzero in the orthorhombic phase where the $C_4$ symmetry of the Fe unit cell
is broken at the structural transition $T_S$. Note that, in certain systems the structural and the magnetic transition are
simultaneous ($T_S$=$T_N$), and, since the magnetic order by itself breaks $C_4$ symmetry, it is not clear whether the nematicity is
a mere by-product of the magnetic order. Consequently, the issue of nematicity is more sharply posed for those systems where the
structural transition precedes the magnetic one ($T_S>T_N$) leaving a finite temperature interval where $C_4$ symmetry is broken
while the system remains paramagnetic \cite{Ni2008,Chu2009,Luetkens2009,Parker2010}.
The extreme example of this trend appears to be FeSe where only a structural transition \cite{Margadonna2008,McQueen2009} is detected
and the system remains paramagnetic
down to its SC phase \cite{Hsu2008,Imai2009}, hinting that nematic degrees of freedom may not be necessarily linked to magnetic ones.

\par
The microscopic origin of the nematic order is currently not known with certainty.
One scenario is that the structural transition is, in fact, an instability driven by
the anharmonic lattice potential, in which case the primary order parameter is the
lattice orthorhombicity, and the electronic degrees of freedom are secondary order
parameters that passively follow the symmetry breaking induced by the lattice strain. A second scenario
is that the $C_4$ symmetry breaking is driven by electronic interactions, in which case
the primary order parameter is electronic in origin. Within this picture
one possibility is the
spin-nematic transition whereby the spins of the two Fe sublattices phase-lock, which breaks $C_4$ symmetry,
without developing a spontaneous magnetization, i.e., without breaking time reversal
symmetry~\cite{Chandra1990,Fang2008,Xu2008,Qi2009,Fernandes2010,Paul2011,Cano2011,Fernandes2012,Fanfarillo2015}. A second possibility
is ferro-orbital ordering \cite{Kruger2009,Lee2009,Chen2010a, Lv2010, Onari2012}, where either the
occupations or the hopping matrix elements (or both) of the $d_{xz}$ and the $d_{yz}$ orbitals of Fe become inequivalent.
Besides these two scenarios, other possibilities include a d-wave Pomeranchuk instability~\cite{Zhai2009} in which the Fermi surfaces
undergo symmetry breaking distortions due to interaction effects,
as well as a valley density wave~\cite{Kang2011}.
\par
On the experimental side, initial studies have focused on the strong anisotropy of the electronic properties in the orthorhombic $C_4$ symmetry
broken phase. Transport \cite{Chu2010,Tanatar2010b,Ying2011, Blomberg2013,Ishida2013,Jiang2013}, optical
conductivity \cite{Dusza,Nakajima2011,Nakajima2012,Mirri2014a,Mirri2014b}, ARPES \cite{Yi2011,Kim2011b} and neutron
scattering \cite{Harriger2011,Lu2014,Lu2015} (reviewed in a separate contribution to this issue \cite{Inosov-CRAS}) performed
on mechanically detwinned crystals all reported considerable electronic anisotropies. While it has been argued that the magnitudes
of the measured anisotropies are too large to be due to the lattice orthorhombicity (which is 0.4 percent at most), such arguments
can be at best quantitative, and therefore they do not convincingly rule out the lattice-driven scenario. Furthermore, even within the
electronic-driven scenario, the above experiments cannot uniquely identify whether the primary order parameter is composed of electronic
charge, spin or orbital degrees of freedom~\cite{Fernandes-Review-2014}.

\par
One difficulty in interpreting the above experiments is that, in the symmetry-broken phase, all the above
order parameters are non-zero, and consequently it is difficult to identify the one which is most relevant.
From this point of view it is desirable to set experiments that measure relevant susceptibilities
in the $C_4$ symmetric phase, and search for signatures of nematic fluctuations that soften upon approaching $T_S$. One obvious
possibility is the measurement of the orthorhombic elastic constant which measures the force constant associated with orthorhombic
strain of the lattice~\cite{Fernandes2010,Goto2011,Yoshizawa2012,Bohmer2014}.
Such studies are reviewed in a separate contribution to this issue \cite{Bohmer-CRAS}. However, note that an
elastic constant measurement cannot, by itself, clearly distinguish between a lattice-driven from an electronic-driven
scenario of nematicity. This is because an elastic constant measurement is a thermodynamic probe, and therefore, once the interaction
between electrons with the acoustic phonons is taken into account, in
both scenarios one would conclude softening of the relevant elastic constant. Next, in the spin-nematic scenario
the fluctuations of the order parameter (which is a two-spin operator) involve a four-spin susceptibility.
This poses a technical difficulty because, while NMR and neutron scattering can give crucial information on the evolution of the spin
fluctuation spectrum \cite{Ning2010,Nakai2010,Fu2012,Zhao2008b,McQueeney2008,Ewings2008,Harriger2011}
and the spin susceptibility at the antiferromagnetic wave-vector, they
cannot easily access the four-spin susceptibility. This means that, at present there is no direct probe to test
the fluctuations associated with spin-nematic order parameter.
An alternative way to probe the nematic susceptibility was pioneered by Chu et al. \cite{Chu2012,Kuo2013,Kuo2014} who were able to
extract the nematic component of the elasto-resistivity tensor in the tetragonal phase. This was achieved by performing strain dependent
measurements of the transport anisotropy. Because the method allows a direct extraction of the lattice-free electronic nematic
susceptibility, the observed divergence is a strong evidence for an electronic-driven structural phase transition. A drawback however,
is the difficulty to associate the extracted nematic susceptibility to a microscopic nematic order parameter.
\par
On the other hand, in this review we argue that electronic Raman scattering allows a direct access to
the fluctuations of the charge nematic order parameter, or equivalently the d-wave Pomeranchuk order parameter.
This ability of Raman measurements has been somewhat overlooked in the past, although earlier Raman
experiments in underdoped cuprates could possibly be interpreted along these lines \cite{Tassini2005}. We
show in this review that Raman experiments in Fe SC give compelling evidences for the presence of
nearly-critical charge nematic fluctuations in the tetragonal phase.
These experiments also allow a direct extraction of the associated nematic susceptibility, which contains
information about the incipient phase instability involving the \emph{purely electronic} degrees of freedom.
This is because the electronic Raman response is a spectroscopic probe, and it is ``opaque'' to the acoustic phonons in the system.
In fact, this property of the electronic Raman response function provides a \emph{qualitative method} to distinguish between
lattice-driven versus electronically-driven scenarios of nematicity. It can be shown that in the former case the extracted nematic
susceptibility from the Raman data should not show any signature of softening with lowering temperature. The fact that
in Fe SC one does see softening \emph{proves conclusively that the nematicity is electron-driven rather than lattice-driven}.
This establishes Raman scattering as a key probe of translation symmetry preserving Fermi surface distortions
which can be used to investigate other correlated electron systems such as the bilayer ruthenates and the
cuprates where this kind of instability has been proposed but not yet confirmed unequivocally from an
experimental point of view.

\par

This review is divided into two main parts, one theoretical and the other experimental.  We start with the theory
part which aims at giving a simple framework to understand how Raman scattering can be used to probe the charge nematic susceptibility
and its associated dynamical fluctuations in an electron system. The approach is quite general, and the main
features are expected to hold for the case of the Fe SC. In \ref{subsec:model} we consider a generic one band model with a charge
nematic or Pomeranchuk instability where the Fermi surface breaks the $C_4$ symmetry.
In \ref{subsec:instability} we provide analytical expressions for its critical fluctuations within the
random-phase approximation (RPA). In \ref{subsec:B1g} we show that electronic Raman scattering directly couples to the charge
nematic fluctuations, provided the appropriate symmetry channel is probed. In the case of the Fe SC this is the  $B_{1g}$ channel
(which transforms as ($x^2-y^2$) in the 1Fe/cell notation).
We show in \ref{subsec:quasi-elastic} that the presence of these nematic fluctuations leads to the
emergence of a quasi-elastic peak whose linewidth tends to vanish at the incipient pure electronic
(ie, one without lattice coupling) nematic phase transition. We
further show in \ref{subsec:lattice} that Raman scattering measurements can access the charge nematic susceptibility, but only in
the dynamical limit as opposed to the static limit which is relevant for the definition of the thermodynamic phase transition
involving the $C_4$ symmetry breaking. While these two limits are the same for a pure electronic system, the coupling to the lattice induces
a key difference between the two limits making Raman scattering measurement essentially blind to the linear
coupling between the electron-nematic variable and the lattice orthorhombic strain. We conclude the theory part by
discussing in \ref{subsec:multiorb} some of the additional
subtleties associated with the multi-band nature of Fe SC. In particular we point out the existence of different flavors of
charge nematicity when the orbital quantum number is taken into account. However, Raman measurements cannot distinguish between
the various charge nematic order parameters that are possible in a multi-orbital environment.

\par
In the experimental part, after reviewing briefly the details of the Raman experiments in \ref{sec:Raman-exp},
we discuss the observation of charge nematic fluctuations in the tetragonal phase of electron-doped Ba(Fe$_{1-x}$Co$_x$As)$_2$.
In \ref{subsec:nematic-CoBa122} we focus on the behavior of the extracted charge nematic susceptibility as a function of Co
electron doping and temperature. We then compare in \ref{subsec:exp-comp} Raman results with two other complementary probes of nematic
fluctuations in the tetragonal state, elastoresistiviy and elastic constant measurements, and conclude that all three measurement
are consistent with an electronic driven structural phase transition. We briefly comment on the role of disorder in \ref{subsec:disorder}
by comparing Ba122 and Sr122 systems. In \ref{subsec:QEP} we discuss the finite frequency spectra of the nematic fluctuations and show
that it is consistent with expectations from a simple mean-field approach
of the nematic phase transition (\ref{subsec:RPA}). We then show in \ref{subsec:Raman-shear} that Raman and shear modulus measurements
can be consistently reproduced using a simple Landau-type mean-field picture with a linear coupling between charge and lattice nematicity.
We conclude this review by addressing the fate of nematic fluctuations in the superconducting state (\ref{sec:nematic-SC}) and show that
they can give arise to a novel collective mode, a nematic resonance,
near the nematic quantum critical point. Experiments on electron doped Fe SC (Co-Ba122 and Co-Na111) appears to support its existence.

\section{Theory: Charge nematic instability}
\label{sec:nematic-theory}

In this section we provide a microscopic description of charge nematic instability triggered by a phenomenologically-introduced
electron-electron interaction.

\subsection{Model}
\label{subsec:model}

We consider a system of interacting electrons on a square lattice, simultaneously scattered by point-like
impurity described by the  Hamiltonian
\beq
\label{eq:ham}
\ham = \ham_0 + \ham_I + V.
\eeq

In the above
\beq
\label{eq:ham0}
\ham_0 = \sum_{\bk, \si} \epsilon_{\bk} c^{\dagger}_{\bk, \si} c_{\bk, \si}
\eeq
is the bare Hamiltonian of a band of electrons with dispersion $\ep_{\bk}$, having lattice momentum $\bk$ and
spin $\si$ as quantum numbers, and described by usual creation/annihilation operators
$(c^{\dagger}_{\bk, \si}, c_{\bk, \si})$. Note that, while the Fe-based superconductors (Fe SC) are multi-band systems, here,
for the sake of simplicity, we restrict ourselves to a one-band model. Our main goal
is to discuss certain qualitative physics, rather than quantitative ones, involving Raman spectroscopy near a
charge nematic transition. We expect the main conclusions of this analysis to remain unchanged for a multi-band environment.
Of course, in a multi-band system there are inter-band transitions that contribute to the Raman response, which are absent
in a one-band model.
However, typically such inter-band transitions are not related to criticality involving
charge nematic transition. The latter is essentially a Fermi surface instability, and consequently, its
critical fluctuations involve only intraband excitations. In Sec.~\ref{subsec:multi-orbital} we comment about
the additional subtleties associated with nematic instabilities in a multi-orbital and multi-band environment.

\begin{figure*}
\begin{center}
\includegraphics[clip,width=0.7\linewidth]{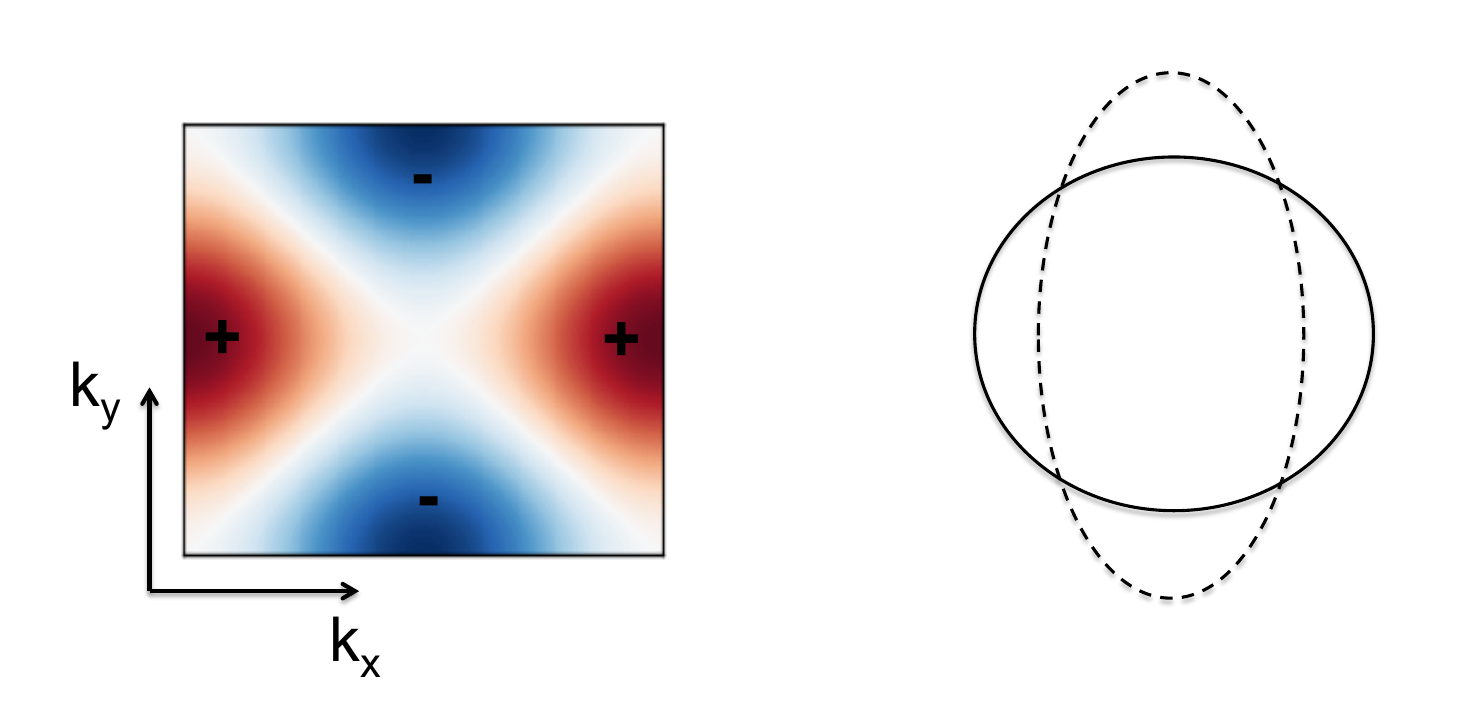}
\caption{Left: $B_{1g}$ symmetry form factor $h_k$ of the Pomeranchuk or charge nematic operator in k-space.
Right: associated Fermi surface deformation in a one-band model. The original spherical Fermi surface is distorted into
an elliptical one in the charge nematic phase.}
\label{fig0a}
\end{center}
\end{figure*}

The interaction between the electrons is described by
\beq
\label{eq:hamI}
\ham_I = - \frac{g_0}{2} \sum_{\bq} O_n(-\bq) O_n (\bq).
\eeq
The operator
\beq
\label{eq:Onq}
O_n(\bq) \equiv \frac{1}{\sqrt{\mathcal{N}}} \sum_{\bk, \si} f_{\bk, \bq} c^{\dagger}_{\bk + \bq, \si} c_{\bk, \si},
\eeq
where $f_{\bk,\bq} = (h_{\bk} + h_{\bk + \bq})/2$ with the $B_{1g}$ (or equivalently $x^2-y^2$) form factor
$h_{\bk} = \cos k_x - \cos k_y$,
is the Fourier transform of the charge nematic operator (see Fig. \ref{fig0a})
\beq
\label{eq:Onr}
O_n (\br_i) = \frac{1}{4} \left[ \left( c^{\dagger}_i c_{i - \hat{x}} + c^{\dagger}_i c_{i + \hat{x}} + {\rm h.c.}
\right) - \hat{x} \rightarrow \hat{y} \right].
\eeq
$\br_i$ is the position of the lattice site $i$, and $\mathcal{N}$ is the total number of sites. In the high
temperature $C_4$-symmetric phase $\langle O_n (\br_i) \rangle =0$, while in the symmetry broken
phase $\langle O_n (\br_i) \rangle = \delta_0$, $\forall i$, implying a charge nematic phase that preserves
translation symmetry but not the discrete $\pi/2$ rotational symmetry, such that the
hopping matrix elements along $\hat{x}$ and $\hat{y}$ are inequivalent. This is the lattice version of a
Pomeranchuk transition \cite{Pomeranchuk,Fradkin-review} where a spherical (or circular in two dimensions) Fermi surface
becomes ellipsoidal (or elliptical), and indeed in the following we do not distinguish between the
names ``charge nematic'' and ``Pomeranchuk''. We take the constant $g_0 > 0$, implying
that $C_4$ symmetry-breaking distortions of the Fermi surface are energetically favored by the interaction term.
This is a pre-requisite for a charge nematic instability, at least in weak coupling theories involving
random phase approximation.

Evidently, there is non-trivial physics involved in transforming the usual short-range
repulsion between the electrons into an interaction given by Eq.~(\ref{eq:hamI}), and in the Fe SC this can be a consequence of
spin, charge, or orbital fluctuations \cite{Fernandes-Review-2014} (The nature of magnetic interactions and their interplay with
orbital degrees of freedom in Fe SC is reviewed in a separate contribution to this issue \cite{Bascones-CRAS}). However, an inquiry
into the origin of this interaction is related to the question as to what triggers the nematic transition. To the best of our knowledge,
this issue is not entirely settled for all the Fe-based systems, and is beyond
the scope of the current review. Instead, here we adopt a more phenomenological point of view whereby, assuming the existence of such a
transition, we write down a minimal effective theory that describes the transition and the critical phenomenon associated with it.
It is in this low-energy effective theory sense that one should understand the above interaction term.

The last term
\beq
\label{eq:hamV}
V = V_0 \sum_{\bk, \bq, \si} c^{\dagger}_{\bk + \bq, \si} c_{\bk, \si},
\eeq
describes scattering of electrons with isotropic point-like impurity potentials.
The role of this term is to provide a finite lifetime to the electrons. In the following the impurity term
does not affect the description of the charge nematic instability. However, the inclusion of finite lifetime
is crucial for a meaningful discussion of the Raman response function.
In a perfect metal, where quasi-particles are infinitely long-lived, the Raman response, involving
intra-band particle-hole excitations and zero momentum transfer with the photons,
vanishes. This is due to the fact that the constraints from energy and momentum conservation cannot be satisfied simultaneously. Adding
impurity scattering is an effective way to bypass the latter, and to take into account the intra-band contribution
that is always, in practice, present.

\subsection{Instability and critical fluctuations}
\label{subsec:instability}

We first discuss the effect of the impurity scattering. We assume that the impurity potential is
weak enough such that it can be treated in Born approximation. This provides a finite lifetime to
the electrons given by
\beq
\tau^{-1} = 2 \pi n_i V_0^2 \rho_0,
\label{eq:tau}
\eeq
where $n_i$ is the impurity concentration, and $\rho_0$ is the density of states at Fermi energy.

Next, we treat the interaction in random phase approximation. The description of the charge nematic
instability is facilitated by introducing the Hubbard-Stratanovich field $\phi_n(\bq)$
to decouple the interaction which can be rewritten as
\beq
\label{eq:hamI-2}
\ham_I = \frac{g_0}{2} \sum_{\bq} \left[ \phi_n(-\bq) \phi_n (\bq)
+  \phi_n(-\bq) O_n(\bq)  \right].
\eeq
The second term above describes interaction between electrons and the Hubbard-Stratanovich field $\phi_n(\bq)$,
which is shown graphically in Fig.~\ref{fig0b} (a).
With this rearrangement the theory is formally quadratic in the fermionic variables, which can
be integrated out. This leads to the action in terms of the critical variable
\beq
\label{eq:action}
S\left[ \phi_n \right] = \sum_{\bq, i \nu_n} \chi^{-1}_n (\bq, i \nu_n)
\left| \phi_n (\bq, i \nu_n) \right|^2,
\eeq
where $\chi_n (\bq, i \nu_n)$ is the nematic susceptibility given by
\beq
\label{eq:chi-n}
\chi^{-1}_n (\bq, i \nu_n) = g_0 \left[ 1 - g_0 \Pi_n(\bq, i \nu_n) \right],
\eeq
with the nematic polarization
\beq
\label{eq:Pi-n}
\Pi_n(\bq, i \nu_n) = - \frac{2}{\beta} \sum_{\om_n, \bk} f^2_{\bk, \bq} G_{\bk}(i \om_n) G_{\bk + \bq}(i \om_n + i \nu_n).
\eeq
In the above $\beta = 1/(k_B T)$, where $k_B$ is the Boltzmann constant, and the factor 2 is due to summation over the
spin index. The electron Green's function is given by
\beq
\label{eq:G-fn}
G_{\bk}(i \om_n)^{-1} = i \om_n - \ep_{\bk} + i/(2 \tau) {\rm Sgn} (\om_n).
\eeq
Note that, since the impurity potential is isotropic, for symmetry reasons the vertex correction due to impurity
scattering affects neither the $\bq \rightarrow 0$ limit of $\chi_n (\bq,0)$, which controls the charge nematic transition
temperature, nor the susceptibility
$\chi_n (0, i \nu_n)$, which will be relevant when we discuss the Raman response in the next section. Consequently,
in the current model the role of the impurity is restricted to providing finite lifetime to the electronic excitations.

It is convenient to divide the frequency and momentum dependencies of the nematic polarization into two parts,
namely the contributions from the high-energy and the low-energy electrons, such that
\beq
\label{eq:Pi-n-2}
\Pi_n(\bq, i \nu_n) = \Pi_n(0, 0) + \Pi_n(\bq, i \nu_n)_{\rm high} + \Pi_n(\bq, i \nu_n)_{\rm low}.
\eeq
Since we do not expect any singular contribution from the high-energy electrons, the resulting frequency and momentum
dependence is analytic. To lowest order we expect
\beq
\label{eq:Pi-n-high}
\Pi_n(\bq, i \nu_n)_{\rm high} =  c_1 (q/k_F)^2 + c_2 (\nu_n/E_F)^2.
\eeq
In the above $\Pi_n(0,0)$ and the coefficients $c_{1,2}$ depend on the details of the band structure. The charge nematic
transition takes place when the Stoner criterion
\beq
\label{eq:r0}
r_0 \equiv 1 - g_0 \Pi_n(0,0) = 0
\eeq
is satisfied. In general $r_0$, which is related to the nematic correlation length $\xi_n$ by
$r_0 = (a/\xi_n)^2$ with $a$ being the unit cell length, is temperature dependent and it decreases
as the nematic correlation length increases upon approaching the instability with lowering temperature.
In electron-doped Ba122 it is now known from elastic constant
measurements (discussed in a separate contribution to this issue \cite{Bohmer-CRAS}) and also from Raman scattering
measurements (described later in this review) that $r_0$ is linear in temperature over a wide range, i.e.
\beq
\label{eq:r0-2}
r_0(T) = \tilde{r}_0 (T - T_0),
\eeq
where $T_0$ is the charge nematic transition temperature.

The low-energy contribution $\Pi_n(\bq, i \nu_n)_{\rm low}$ determines the dynamical properties of the charge nematic fluctuations,
and its evaluation is quite standard. The $\ep_{\bk}$-integral can be performed
by linearizing the electronic dispersion. This gives to leading order
\[
\Pi_n(\bq, i \nu_n)_{\rm low} = - i \nu_n \rho_0 \int_0^{2\pi} \frac{d \theta_{\bk}}{2 \pi}
\frac{h_{\bk}^2}{i S_{\nu_n} - v_F q \cos (\theta_{\bk} - \theta_{\bq})}.
\]
In the above $S_{\nu_n} = \nu_n + {\rm Sgn} (\nu_n)/\tau$, $v_F$ is Fermi velocity, and $\theta_{\bk}$ is the
angle of $\bk$ measured from one of the two equivalent major axes of the unit cell. The angular integral can be
performed analytically if we approximate the $B_{1g}$ form factor by $h_{\bk} \approx - \cos (2 \theta_{\bk})$.
From the angular integral we get
\begin{align}
\label{eq:Pi-n-low}
\Pi_n(\bq, i \nu_n)_{\rm low} &=  \frac{i \nu_n \rho_0 }{v_F q}
\left[ \cos^2 (2 \theta_{\bq}) I_C ( a_{\bq,\nu_n}) \right. \nonumber \\
&+ \left. \sin^2 (2 \theta_{\bq}) I_S ( a_{\bq,\nu_n}) \right],
\end{align}
where
$a_{\bq,\nu_n} = [\nu_n + {\rm Sgn}(\nu_n)/\tau]/(v_F q)$,
\begin{align}
\label{eq:I-C}
I_C(a) & \equiv - \int_0^{2\pi} \frac{d \theta}{2 \pi} \frac{\cos^2 2 \theta}{ia - \cos \theta}
\nonumber \\
& = i (1+ 2a^2) \left[ \frac{(1+ 2a^2)}{\sqrt{1+a^2}} {\rm Sgn}(a) - 2a \right],
\end{align}
and
\begin{align}
\label{eq:I-S}
I_S(a) & \equiv - \int_0^{2\pi} \frac{d \theta}{2 \pi} \frac{\sin^2 2 \theta}{ia - \cos \theta}
\nonumber \\
& = 2ai \left[ 1+ 2a^2 - 2 \left| a \right| \sqrt{1+a^2} \right].
\end{align}Thus, the overall $\bq$-dependence of the nematic polarization is
anisotropic, which is eventually a consequence of the form factor associated
with the nematic variable defined in Eq.~(\ref{eq:Onq}) \cite{Zacharias2009}.
In the above the ratio $a_{\bq,\nu_n}$ is large for temporal fluctuations and it is small
for spatial fluctuations.

\emph{Quasi-static limit, $a_{\bq,\nu_n} \ll 1$}:
This limit is relevant for studying the thermodynamic signatures of the charge nematic instability. Using the
properties
\begin{align}
I_C(a \rightarrow 0) & = i {\rm Sgn}(a) - 2ia,
\nonumber \\
I_S(a \rightarrow 0) &= 2ia,
\nonumber
\end{align}
we get in this limit
\begin{align}
\label{eq:chi-n-static}
\chi^{-1}_n (\bq, i \nu_n) &= g_0 \left[ r_0 + c_1 \frac{q^2}{k_F^2}
+ c_2 \frac{\nu_n^2}{E_F^2}
+ 2 c_3 \frac{\left| \nu_n \right|}{v_F q} \cos^2 2 \theta_{\bq} \right.
\nonumber \\
& - \left.
4 c_3  \frac{\nu_n^2}{(v_F q)^2} ( 1 + (\left| \nu_n \right| \tau)^{-1}) \cos 4 \theta_{\bq} \right],
\end{align}
with $c_3 = g_0 \rho_0$.

\emph{Quasi-dynamical limit, $a_{\bq,\nu_n} \gg 1$}:
This limit is relevant for studying the signatures of the critical mode
$\phi_n(\bq)$ in Raman spectroscopy \cite{Kontani2014}. This is because in typical Raman scattering experiments the
momentum $\bq$ given to the electrons by the visible photons is negligible compared to the Fermi momentum. Using
\[
I_C(a \rightarrow \infty) = I_S(a \rightarrow \infty) = i/(2a),
\]
we get in this limit
\begin{align}
\label{eq:chi-n-dynamic}
\chi^{-1}_n (\bq, i \nu_n) = g_0 \left[ r_0 + c_1 \frac{q^2}{k_F^2}
+ c_2 \frac{\nu_n^2}{E_F^2}
+ c_3 \frac{\left| \nu_n \right|}{\left| \nu_n \right| + 1/\tau} \right].
\end{align}
In this limit the momentum anisotropy is absent from the leading order terms.

\emph{Analyticity of $\chi_n(0, 0)$}:
From Eqs.~(\ref{eq:chi-n-static}) and (\ref{eq:chi-n-dynamic}) it is important to note that
$\chi_n(0,0)$ obtained from the static and the dynamic limits are the same, i.e.,
for an infinitesimal $\eta > 0$
\beq
\label{eq:chi-n-limit}
\lim_{q \rightarrow 0} \lim_{\om \rightarrow 0} \chi_n (\bq, \om + i \eta)
= \lim_{\om \rightarrow 0} \lim_{q \rightarrow 0} \chi_n (\bq, \om + i \eta)
= \frac{1}{g_0 r_0}.
\eeq
This implies that $\chi_n(\bq, \om + i \eta)$ is analytic at zero momentum and frequency.

As a quick aside, the above behavior is to be contrasted with the susceptibility of a conserved
quantity such as the charge susceptibility, which as we will show below is not what is measured in the Raman response. The charge susceptibility is defined by
\beq
\label{eq:chi-c}
\chi_c (\bq, i \nu_n) \equiv \int_0^{\beta} \langle T_{\tau} \rho_{-\bq}(\tau)
\rho_{\bq}(0) \rangle e^{i \nu_n \tau},
\eeq
where $T_{\tau}$ is the imaginary time ordering operator, and
\beq
\label{eq:rho}
\rho_{\bq} \equiv \sum_{\bk, \si}  c^{\dagger}_{\bk + \bq, \si} c_{\bk, \si},
\eeq
is the Fourier component of the charge density operator. In this case it is well
known that the uniform charge susceptibility from the static limit is finite with
\beq
\label{chi-c-static}
\lim_{\bq \rightarrow 0} \chi_c(\bq, 0) = \chi_c \neq 0,
\eeq
where $\chi_c$ is the charge compressibility of the electronic system. On the other hand
the dynamical limit vanishes with
\beq
\label{eq:chi-c-dynamic}
\chi_c(0, \om + i \eta) = 0.
\eeq
This is a consequence of particle number conservation, i.e., $\left[ \ham , \rho_{\bq=0} \right] = 0$,
which in turn follows from the global $U(1)$ symmetry of the Hamiltonian. In other words, this
symmetry enforces a non-analyticity at zero frequency and momentum, and
\beq
\label{eq:chi-c-limit}
\lim_{q \rightarrow 0} \lim_{\om \rightarrow 0} \chi_c (\bq, \om + i \eta)
\neq \lim_{\om \rightarrow 0} \lim_{q \rightarrow 0} \chi_c (\bq, \om + i \eta).
\eeq

By contrast, the uniform nematic operator is not a conserved quantity,
since $\left[ \ham_I , O_n(\bq=0) \right] \neq 0$, and therefore there is
no physical reason to expect a similar non-analyticity in $\chi_n (\bq, \om + i \eta)$
at zero frequency and momentum in \emph{purely electronic models}, i.e., those where the coupling of the electrons
to the lattice strains is ignored. As we discuss in
the next section, the analyticity implied by Eq.~(\ref{eq:chi-n-limit}) is important
for interpreting the signature of the charge nematic instability in Raman response.

Note that, in order to establish the analyticity of $\chi_n(0, 0)$ it is crucial to consider
electrons with finite lifetime. It is easy to check that Eq.~(\ref{eq:chi-n-limit}) does not
hold for an ideal metal for which $\tau^{-1} \rightarrow 0$. However, such a non-analyticity,
which is not associated with any symmetry, and whose origin can be traced to the fact that for
an ideal metal the phase space for particle-hole excitations is sharply defined, is rather
an artefact. In practice, the electrons have a finite lifetime, and this ensures that the
phase space for particle-hole excitations is no longer sharply defined.

\section{Theory: Electronic Raman response near charge nematic instability}
\label{sec:raman-theory}

In this section we discuss the characteristic signatures of a charge nematic instability
in electronic Raman spectroscopy. We also discuss how electron-lattice coupling affects
such an instability. We argue that the Raman response is ``blind'' to this coupling,
and therefore it is an \emph{ideal tool for studying the bare electronic nematic correlations}.

\subsection{$B_{1g}$ response, static and dynamic limits}
\label{subsec:B1g}

\begin{figure*}
\begin{center}
\includegraphics[width=17cm,trim=0 0 0 0]{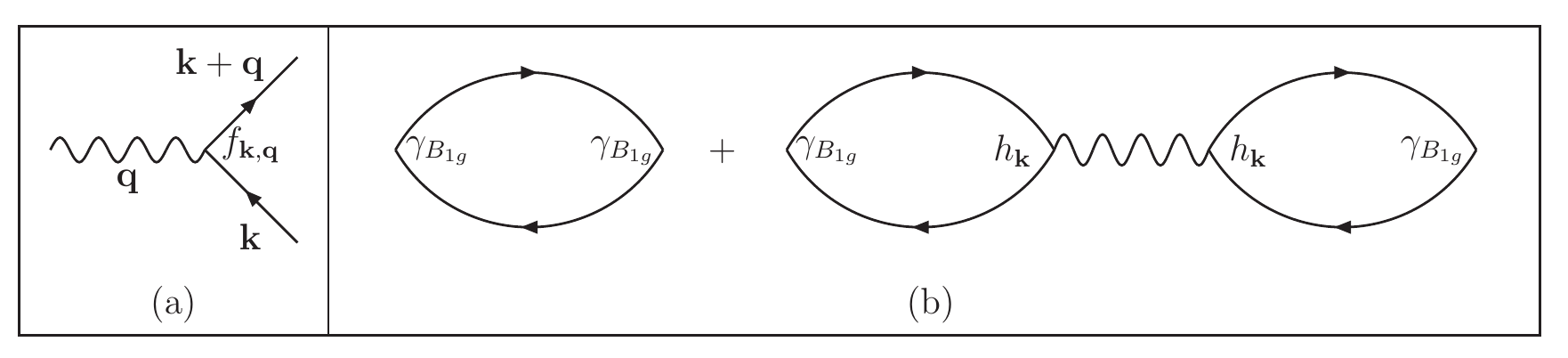}
\caption{
(a) Interaction between electrons (solid lines) and the charge nematic boson (wavy line)
$\phi_n$. The interaction vertex has nontrivial momentum dependence given by
$f_{\bk,\bq} = (h_{\bk} + h_{\bk + \bq})/2$ with the $B_{1g}$ form factor
$h_{\bk} = \cos k_x - \cos k_y$. (b) Graphs for the Raman response function. The first
term is the quasiparticle contribution which is non-critical. The second term is the
contribution of the nematic boson which has information about criticality associated with the
nematic instability. This contribution is non-zero only in the $B_{1g}$ Raman channel.
}
\label{fig0b}
\end{center}
\end{figure*}

The theory underlying the electronic Raman spectroscopy for correlated systems has been reviewed
elsewhere \cite{Klein,Devereaux2007}. Here, we simply remind few salient points associated with this experimental
technique. Accordingly, we define
the stress tensor by
\beq
\label{eq:stress-tensor}
T_{\al \be}(\bq) \equiv \sum_{\bk, \si} \frac{\ptl^2 \ep_{\bk}}{\ptl k_{\al} \ptl k_{\be}}
c^{\dagger}_{\bk + \bq/2, \si} c_{\bk - \bq/2, \si},
\eeq
and the associated correlator as
\beq
\label{eq:stress-correlator}
\chi_{\al \be} (\bq, i \Om_n) \equiv \int_0^{\be} d \tau \langle T_{\tau} T_{\al \be}^{\dagger}(\bq, \tau)
T_{\al \be}(\bq, 0) \rangle e^{i \Om_n \tau}.
\eeq
Following Kubo formalism, the analytic continuation of the above gives the response function
\begin{align}
\label{eq:stress-response}
\chi_{\al \be} (\bq,  \Om) &= \lim_{i \Om_n \rightarrow \Om + i \eta} \chi_{\al \be} (\bq, i \Om_n)
\nonumber \\
&= i \int_0^{\infty} dt e^{i \Om t} \langle \left[ T_{\al \be}^{\dagger}(\bq, t) \ , \ T_{\al \be}(\bq, 0)
\right] \rangle.
\end{align}
The imaginary part of the stress tensor response function at its dynamical limit, i.e.,
$\chi^{\prime \prime}_{\al \be} (\bq=0, \Om) \equiv {\rm Im} \chi_{\al \be} (\bq =0,  \Om)$, is accessible
by means of Raman scattering experiment. This is because the associated scattering cross-section is proportional to the
correlation function $S_{\al \be}(\Om) \equiv \langle T_{\al \be}^{\dagger} (\Om) T_{\al \be} (\Om) \rangle$,
which in turn satisfies the fluctuation-dissipation theorem
\beq
\label{eq:fluc-diss}
S_{\al \be}(\Om) = \frac{1}{\pi} \left[ 1 + n_B(\Om) \right] \chi^{\prime \prime}_{\al \be} (\bq=0, \Om),
\eeq
where $n_B(\Om)$ is the Bose function.

In the Fe SC, using the notations of a unit cell with 1Fe/cell, the quantity of interest is
the $B_{1g}$ stress tensor given by
\beq
\label{eq:Big-stress}
T_{B_{1g}}(\bq) \equiv T_{xx}(\bq) - T_{yy}(\bq).
\eeq
Within the effective mass approximation \cite{Ashcroft}, valid for non-resonant Raman scattering, we define the associated
Raman vertex as
\beq
\label{eq:gamma-k}
\gamma_{B_{1g}}(\bk)  \equiv \left( \ptl_{k_x}^2 - \ptl_{k_y}^2 \right) \ep_{\bk}
= 2t_1 h_{\bk},
\eeq
where the dispersion of Eq.~(\ref{eq:ham0}) is given by $\ep_{\bk} = -2t_1 (\cos k_x + \cos k_y) + \cdots$, with
$t_1$ being the nearest neighbor hopping parameter. In principle, $\gamma_{B_{1g}}(\bk)$ can include higher
harmonics of the same $B_{1g}$ symmetry, such as that coming from the third nearest neighbor hopping, if present in $\ham_0$.
In the following we ignore such terms since they do not affect the results qualitatively.

The computation of the $B_{1g}$ correlator within random phase approximation (RPA) is quite straightforward.
We get
\beq
\label{eq:R-B1g-1}
\chi_{B_{1g}} (\bq, i \Om_n) = \Pi_{\ga \ga} (\bq, i \Om_n) + g_0^2 \Pi_{\ga h}^2 (\bq, i \Om_n) \chi_n (\bq, i \Om_n),
\eeq
where $\Pi_{\ga \ga}$ is defined like $\Pi_n$ in Eq.~(\ref{eq:Pi-n}) with the form factors $f_{\bk,\bq}^2$ replaced by
$\gamma_{B_{1g}}^2(\bk)$, and $\Pi_{\ga h}$ is defined similarly with the form factors $\gamma_{B_{1g}}(\bk) h_{\bk}$.
The graphical representation of these two terms is shown in Fig.~\ref{fig0b} (b). Note that, the second term, which
contains the critical contribution, is non-zero only in the Raman $B_{1g}$ channel. In the $A_{1g}$ and the $B_{2g}$
channels this term is zero by symmetry. Indeed, in experiments the signature of the charge nematic
instability is observed only in the $B_{1g}$ Raman channel.
Since, $\Pi_{\ga \ga} = 4t_1^2 \Pi_n$, and $\Pi_{\ga h} = 2t_1 \Pi_n$, we get, at low frequency and momentum
\beq
\label{eq:R-B1g-2}
\chi_{B_{1g}} (\bq, i \Om_n) \approx 4t_1^2 \chi_n (\bq, i \Om_n).
\eeq
The above proportionality implies that the properties of $\chi_n (\bq, i \Om_n)$, discussed in Sec.~\ref{sec:nematic-theory},
are also relevant for $\chi_{B_{1g}} (\bq, i \Om_n)$ (see also \cite{Yamase2011,Yamase2013}). In particular,
in \emph{purely electronic models}
the $B_{1g}$ response function is analytic at zero frequency and momentum.

In electronic Raman spectroscopy in the $B_{1g}$ geometry we measure a quantity proportional to
$\chi^{\prime \prime}_{B_{1g}} (\bq=0, \Om) \equiv {\rm Im} \chi_{B_{1g}} (\bq =0,  \Om)$.
This quantity can be used
to deduce the frequency-integrated spectral weight of the associated Raman conductivity
$\chi^{\prime \prime}_{B_{1g}} (\bq=0, \Om)/\Om$
which, by Kramers-Kronig relation, gives the real part of the uniform response function. Thus,
\begin{align}
\label{eq:Kramers}
\chi^{\rm dynamic}_{B_{1g}} & \equiv \frac{2}{\pi} \int_0^{\infty} \frac{d \Om}{\Om}
\chi^{\prime \prime}_{B_{1g}} (\bq=0, \Om)
\nonumber \\
& = \lim_{\Om \rightarrow 0} \chi_{B_{1g}} (\bq =0,  \Om).
\end{align}

It is important to distinguish the two conceptually distinct quantities $\chi^{\rm dynamic}_{B_{1g}}$ and
\beq
\label{eq:R-static}
\chi^{\rm static}_{B_{1g}}  \equiv \lim_{\bq \rightarrow 0} \chi_{B_{1g}} (\bq ,  \Om =0).
\eeq
Note that, the thermodynamic instability at a charge nematic transition is associated with a divergence in the
latter quantity. However, since $\chi_{B_{1g}} (\bq, \Om)$ is analytic at zero frequency and momentum in purely
electronic systems, we get
\beq
\label{eq:R-static-dynamic}
\chi^{\rm dynamic}_{B_{1g}}  = \chi^{\rm static}_{B_{1g}} ,
\eeq
and, therefore, the divergence is ``visible'' from the dynamical limit. In other words, one can obtain information
about the divergence of a susceptibility associated with the charge nematic instability from Raman spectroscopy.

\subsection{Signature of instability: quasi-elastic peak}
\label{subsec:quasi-elastic}

In the following we study the low frequency properties of the Raman response near the transition.
Using Eqs.~(\ref{eq:chi-n-dynamic}) and (\ref{eq:R-B1g-2}), for $ \Om \lesssim 1/\tau \ll E_F$ we get
\beq
\label{eq:R-low}
\chi_{B_{1g}} (\bq=0, \Om) = A_0 \left[ r_0 + c_3 \frac{\Om}{\Om + i/\tau} \right]^{-1},
\eeq
with $A_0 = 4t_1^2/g_0$ and $c_3 = g_0 \rho_0$. This implies a Raman response with
\beq
\label{eq:R-low-Im}
\chi_{B_{1g}}^{\prime \prime} (\bq=0, \Om) = \frac{A_0 c_3 \tau^{-1}}{(r_0 + c_3)^2} \left(
\frac{\Om}{\Om^2 + \Gamma^2} \right),
\eeq
where
\beq
\label{eq:Gamma}
\Gamma = \frac{r_0}{(r_0 + c_3) \tau}.
\eeq
Thus, the characteristic signature of the transition in the low frequency
Raman conductivity is a quasi-elastic peak with Lorentzian lineshape that sharpens as the system approaches the
transition, since the  width $\Gamma \rightarrow 0$.

It is useful to note that the $\chi^{\rm dynamic}_{B_{1g}}$ deduced from the experimental data \cite{Gallais2013}
(described in \ref{sec:nematic-122}) has the form
\beq
\label{eq:R-dynamic-expt}
\chi^{\rm dynamic}_{B_{1g}}  = \frac{A_0}{r_0} + B,
\eeq
where $B$ is a non-singular part that is often temperature independent. This apparent violation of the
Kramers-Kronig relation is partly due to the fact that, in practice, the upper cutoff of the
frequency integral of Eq.~(\ref{eq:Kramers}) is finite and is set to a value beyond which the measured
Raman spectra is temperature independent. A second reason for the $B$-term is that part of the electronic Raman
signal observed is symmetry independent and therefore unrelated to criticality.

Finally, we note that, besides $\chi^{\rm dynamic}_{B_{1g}}$ and the width $\Gamma$, a third quantity which is
experimentally accessible is the slope of the Raman response at zero frequency
\begin{align}
\label{eq:S_l}
S_l \equiv \left[ \ptl_{\Om} \chi_{B_{1g}}^{\prime \prime} (\bq=0, \Om) \right]_{\Om=0}
 = \frac{A_0 c_3 \tau}{r_0^2},
\end{align}
where the equality is the result of the random phase approximation. Using $\Gamma \approx r_0/(c_3 \tau)$
close enough to the phase transition, we get a scaling relation between the three experimentally accessible
quantities
\beq
\label{eq:scaling}
\chi^{\rm dynamic}_{B_{1g}}  = \Gamma S_l,
\eeq
provided we ignore the non-singular $B$-term in Eq.~(\ref{eq:R-dynamic-expt}). This provides an independent
check for the validity of the random phase approximation theory.

\subsection{Effects of coupling to the lattice}
\label{subsec:lattice}

A crucial ingredient in the above discussion is the analyticity of the nematic susceptibility
$\chi_n(\bq, i\Om_n)$, and therefore that of the correlator $\chi_{B_{1g}}(\bq, i\Om_n)$ involving
the $B_{1g}$ stress tensor, at zero momentum and frequency. It is this property that allows us
to conclude that the thermodynamic divergence at the charge nematic phase transition, which
shows up in the susceptibility taken to its static limit, is also ``visible'' from the dynamic limit
which is accessible via Raman spectroscopy. As we discussed earlier, in order to demonstrate the
analyticity,
it is important to consider electrons having a finite, frequency independent, lifetime
which, in simplest models, can be typically attributed to impurity scattering.

The above conclusion, however, holds only if we ignore the symmetry-allowed coupling of the electronic charge nematic
operator $O_n(\bq)$ of Eq.~(\ref{eq:Onq}) to the orthorhombic strain of the underlying lattice.
In practice, however, such electron-lattice coupling is always present in crystalline solids. In the Fe SC, the fact
that the $C_4$ symmetry breaking (ignoring the positions of the As and the chalcogen atoms)
is invariably accompanied by orthorhombic distortion, provides phenomenological proof of the presence of such a coupling.
Consequently, it is worthwhile to examine how the above statements concerning the charge nematic transition and its
Raman ``visibility'' are modified once the coupling to the elastic strain is taken into account.

In the following we consider a two dimensional square lattice whose elastic free energy, to lowest order in the
strains, is given by
\beq
\label{eq:lattice}
F_E = \frac{C_{11}}{2} \left( \ep_{xx}^2 + \ep_{yy}^2 \right) + \frac{C_{66}}{2} \ep_{xy}^2
+ C_{12} \ep_{xx}\ep_{yy}.
\eeq
Here $\ep_{ij} \equiv (\partial_i u_j + \partial_j u_i)/2$, with $(i,j) = (x, y)$ are the strains, the vector
${\bf u}$ denotes displacement from equilibrium, and $C_{11}$ etc.\ denote elastic constants in Voigt notation.

In terms of the above the orthorhombic strain is given by $\ep_S(\br) \equiv \ep_{xx}(\br) - \ep_{yy}(\br)$,
and the accompanying bare elastic constant is $C_S^0 \equiv (C_{11} - C_{12})/2$.
We write the symmetry-allowed coupling between the orthorhombic strain and the electronic charge nematic operator
$O_n(\bq)$ as
\beq
\label{eq:el-lattice}
\ham_{\rm el-lattice} = \lambda_0 \sum_{\bq} O_n(\bq) \ep_S (\bq),
\eeq
where $\ep_S (\bq)$ is the Fourier transform of $\ep_S(\br)$, and $\lambda_0$ is the coupling constant
having the dimension of energy.

Studying the detailed implications of the above coupling on an electronic nematic phase transition
is beyond the scope of the current review, and will be presented elsewhere. Here we focus only on the
following salient points that are relevant to the current discussion.

(i) One effect of the coupling is to increase the temperature of the $C_4$ symmetry breaking nematic/orthorhombic
transition from $T_0$, defined in Eq.~(\ref{eq:r0-2}), to
\beq
\label{eq:Tn}
T_S = T_0 + \lambda_0^2/(C_S^0 \tilde{r}_0 g_0).
\eeq
Below $T_S$
the $C_4$ symmetry breaking is manifested both in the electronic sector, where the dispersions becomes $C_2$ symmetric,
as well as in the lattice sector with orthorhombicity $\ep_S \neq 0$. Note that the effective orthorhombic elastic
constant can be expressed as (for a derivation see, e.g., Ref.~\cite{Bohmer-CRAS})
\beq
\label{eq:Cs}
C_S = C_S^0 - \lambda_0^2 \chi_n(0,0),
\eeq
where $\chi_n(\bq, \omega)$ is the bare electronic nematic susceptibility defined in Eq.~(\ref{eq:chi-n}).
This implies that $C_S$ vanishes at $T= T_S$,
and consequently the transition can be detected in experiments
that measure $C_S$ either directly by ultrasound or indirectly by bending techniques.

(ii) At $T_S$ only the effective orthorhombic elastic constant $C_S$ vanishes, while the remaining elastic constants
stay finite. An important consequence of this is that the critical fluctuations are restricted to two high-symmetry
lines $q_x = \pm q_y$ in the two-dimensional Brillouin zone~\cite{Cano2010,Zacharias2015}.
This can be understood from the following. Writing
$q_{1} = (q_x + q_y)/\sqrt{2}$, it can be shown
that along the lines $q_x =  q_y$ and for the polarizations ${\bf n}_1 = (1, -1)$  the acoustic phonon dispersion is
given by $\Om_{1,\bq} = (C_S/\rho)^{1/2} q_1$, where $\rho$ is the atomic mass density. Similarly, along
$q_x = - q_y$ and for the polarization ${\bf n}_2 = (1, 1)$ the dispersion is $\Om_{2,\bq} = (C_S/\rho)^{1/2} q_2$,
where $q_{2} = (q_x - q_y)/\sqrt{2}$.
It can be shown that these are the only two directions in the Brillouin zone for which the phonon velocity
vanishes at $T_S$. This is
because for all other $\bq$ the acoustic phonons excite not just the critical strain $\ep_S$, but also the non-critical
ones whose elastic constants remain finite at $T_S$.

(iii) Finally, and most importantly for the current discussion, as a result of the electron-lattice coupling
the nematic susceptibility acquires a non-analytic correction. Denoting the dressed susceptibility as
$\bar{\chi}_n$, for $\bq = q_{\al} \hat{q}_{\al}$ (summation not implied) with $\al = (1, 2)$, we get that
\beq
\label{eq:chi-n-bar}
\bar{\chi}_n^{-1}(q_{\al} \hat{q}_{\al}, i \Om_n) = \chi_n^{-1}( q_{\al} \hat{q}_{\al}, i \Om_n)
- \frac{\lambda_0^2 q_{\al}^2}{C_S^0 q_{\al}^2 + \Om_n^2}.
\eeq
Here $\hat{q}_{\al}$ are the unit vectors along the two critical directions within the Brillouin zone.
The non-analyticity of the second term is a consequence of translation symmetry. Since moving all the atomic positions by a fixed
displacement does not change the overall energy of the system, the electron phonon coupling, which is the
numerator of the second term above, vanishes in the uniform limit $q_{\al} \rightarrow 0$. As a consequence, the effect of the
acoustic phonons is entirely absent in the dynamical limit and $\bar{\chi}_n(\bq =0, \Om) = \chi_n((\bq =0, \Om))$ \footnote{The in-plane wave-vector transferred is indeed very small in actual Raman experiments, typically $q\sim$ 6.10$^{-3}$ nm$^{-1}$, making finite $\textbf{q}$ effects unobservable in the energy range probed experimentally. The out-of-plane wave-vector transferred can be significantly larger due to the finite penetration depth of the incoming visible photons. We note however the out-of-plane component of $\textbf{q}$ is irrelevant for the coupling to the orthorhombic strain which is purely in-plane (see Eq. \ref{eq:chi-n-bar}.}.
In other words, the Raman response, being opaque to the acoustic phonons, measures the \emph{bare electronic nematicity} and
$\chi^{\rm dynamic}_{B_{1g}} \propto (T- T_0)^{-1}$ deduced from it tends to diverge at the purely electronic temperature scale
$T_0$ \cite{Kontani2014}.
This is in contrast to the inverse of the orthorhombic elastic constant $C_S^{-1} \propto (T- T_S)^{-1}$.
However, the divergence of $\chi^{\rm dynamic}_{B_{1g}}$ is invariably cutoff at the actual $C_4$ symmetry breaking transition $T_S$.
Conversely, since the
electronic Raman response function is unaffected by the acoustic phonons, any \emph{signature of nematicity seen in Raman is
unambiguous proof that it is electronic in origin}.

\subsection{Extension to multi-orbital systems}\label{subsec:multi-orbital}
\label{subsec:multiorb}
\begin{figure}
\centering
\includegraphics[clip,width=0.99\linewidth]{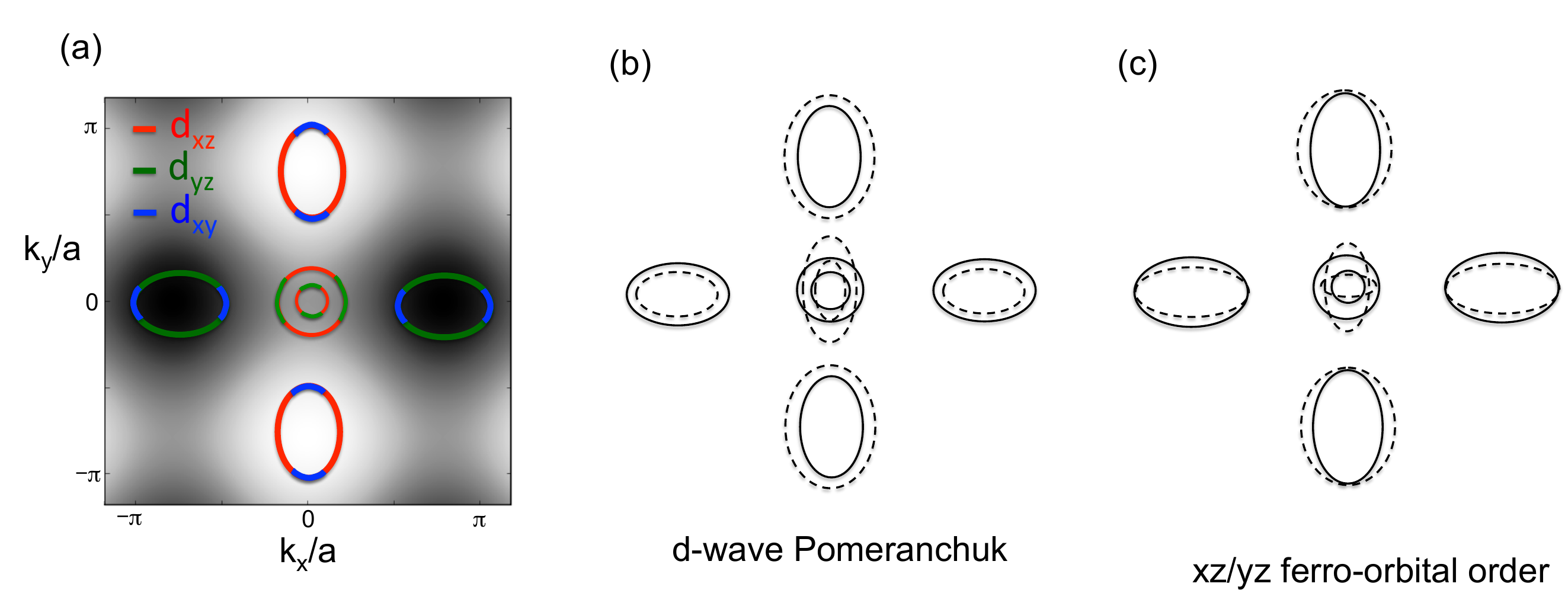}
\caption{ (a) Typical Fermi surface of most Fe SC within a 3 orbital tight binding model
(see e. g. Daghofer et al. \cite{Daghofer2010}). The orbital content of each pocket is indicated. The $B_{1g}$ form factor
$\cos k_x$-$\cos k_y$ is depicted in the background. (b) Fermi surface deformation associated to the $d$-wave Pomeranchuk order $O_{o}^d$  (c) Fermi surface deformation associated with $xz$/$yz$ ferro-orbital order $O_{o}^s$ with $g_k$=1 \cite{Kruger2009, Lee2009}.}
\label{fig1}
\end{figure}

Until now the qualitative physics was described in terms of a single band system in order to simplify the discussion. Nevertheless,
it applies equally well to multi-orbital (and multi-band) systems like the Fe SC. In a multi-orbital environment the main additional
novelty is that, due to the presence of the orbital quantum number, one can construct different flavors of charge nematic order parameters.
The most direct
extension of the single band case described in Eq.~(\ref{eq:Onq}) is a multi-orbital version of the $d$-wave
Pomeranchuk instability where the orbitally resolved electron densities have all $B_{1g}$ nematic form factors :
\beq
\label{eq:PI}
O_{o}^d \equiv \frac{1}{\sqrt{\mathcal{N}}} \sum_{m, \bk} h_{\bk} n_{\bk}^m,
\eeq
where $n_{\bk}$ is the electron density operator, $m$ is the orbital index, and $h_{\bk} = \cos k_x - \cos k_y$. While little
discussed initially this instability has recently been put forward as a candidate order parameter for the orthorhombic phase of
FeSe \cite{Su2015,Jiang2015,Watson2015}.
A second possibility is to define a orbital dependent order parameter of the form :
\beq
\label{eq:OO}
O_{o}^s \equiv \frac{1}{\sqrt{\mathcal{N}}} \sum_{\bk} g_{\bk} (n_k^{xz}-n_k^{yz}),
\eeq
involving the $xz$ and the $yz$ orbitals. Here $g_{\bk}$ is a function with $A_{1g}$ symmetry such as
$g_{\bk} = 1$ (equivalent to $xz$/$yz$ ferro-orbital order \cite{Kruger2009,Lee2009,Lv2009,Chen2010a}), or
$g_{\bk} = \cos k_x + \cos k_y$. Note that, the actual deformations of the Fermi pockets
in the nematic phase, and the axis along which a pocket will elongate/contract, can vary depending on the choice of $g_{\bk}$. In a
two-orbital model relevant for Fe SC the deformations obtained with $g_{\bk} = 1$ are qualitatively similar to that obtained with a
non-zero $d$-wave Pomeranchuk order parameter $O_{o}^d$. One reason for this is that the electron pockets composed mostly
from the $xz$ and the $yz$ orbitals are centered around $(\pi, 0)$ and $(0, \pi)$ respectively, and the form factor $h_{\bk}$ is
approximately a constant, with opposite signs for these pockets. Consequently, the projections of these two order parameters on
the electron
pockets are indistinguishable. The various Fermi surface deformations associated with different choices of the
nematic order parameter are illustrated in Fig.~\ref{fig1}.
\par
\begin{figure}
\centering
\includegraphics[clip,width=0.99\linewidth]{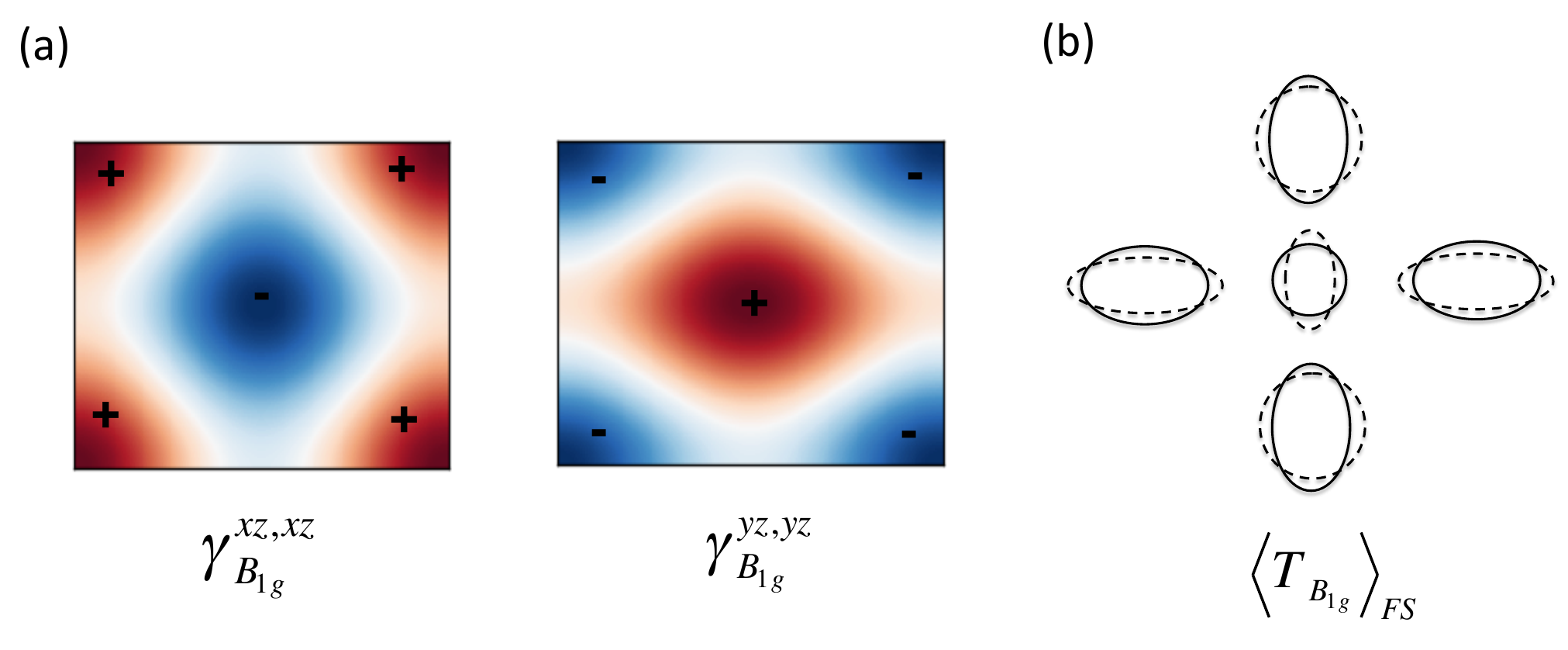}
\caption{(a) k-space structure of the $B_{1g}$  Raman vertex matrix elements in the orbital basis for the two orbital model of
Raghu et al. \cite{Raghu2008} (see also \cite{Belen-vertex2013} for a more realistic 5-band calculation of the Raman vertex).
In this simplified model the main features of the Fermi surface topology of Fe SC can be reproduced with $t_2$=-1.3$t_1$. (b) Fermi
surface deformation associated with the corresponding $B_{1g}$ average stress tensor projected at the Fermi wave-vectors within the
2 orbital model.}
\label{fig1b}
\end{figure}

Overall, we notice that in a multi-orbital system the nematic order parameter can have non-trivial
structure in the orbital space, and it can also be accompanied by momentum space structures that are different than what is possible
in the single-band case described by Eq.~(\ref{eq:Onq}).
From the point of view of Raman spectroscopy it is important to note that, irrespective of the details of their
momentum and the orbital space structures, as long as the nematic order parameter transforms as a $B_{1g}$ object, the
critical fluctuations will be observable in the electronic Raman $B_{1g}$ channel.
This can be illustrated by calculating the form of the $B_{1g}$ stress tensor or vertex for a
specific tight binding model of the Fe SC. In a multi-orbital system one can generalize Eq.~(\ref{eq:gamma-k})
and write the component of the non-resonant $B_{1g}$ Raman vertex in the effective mass approximation \cite{Belen-vertex2013}
\begin{equation}
\gamma^{mn}_{B_{1g}} \equiv \left( \ptl_{k_x}^2 - \ptl_{k_y}^2 \right) \ep^{mn}_{\bk},
\label{multi-vertex}
\end{equation}
where $\epsilon_k^{mn}$ are the components of the tight binding dispersion in the orbital basis. In the minimal
two-orbital ($m,n$=$d_{xz} , d_{yz}$) model of the Fe SC of Raghu et al.  \cite{Raghu2008}, the $B_{1g}$ vertex is
diagonal in orbital space with
\begin{equation}
\label{eq:Raghu-B1g-1}
\gamma_{B_{1g}}\left(\mathbf{k}\right)=\begin{pmatrix}
2(t_1\cos k_x-t_2\cos k_y) & 0  \\
0 & 2(t_2\cos k_x-t_1\cos k_y),  \\
\end{pmatrix}
\end{equation}
where $t_1$ and $t_2$ are the near-neighbor hopping parameters for $\sigma$ and $\pi$ type Fe orbital overlap respectively.
We note that next nearest-neighbor hopping integral along the diagonals of the Fe square plane do not contribute to the $B_{1g}$
Raman vertex. The k-space structure of the two diagonal Raman vertex matrix elements and the associated Fermi surface deformation is
illustrated in Fig. \ref{fig1b} for the tight binding parameters of Raghu et al. \cite{Raghu2008}
\par
In this model the $B_{1g}$ stress tensor is
\beq
\label{eq:Raghu-B1g-2}
T_{B_{1g}}(\bq =0) =  (t_1 + t_2) O_{o}^d + (t_1 - t_2) O_{o}^s,
\eeq
with $g_{\bk} = \cos k_x + \cos k_y$. The $B_{1g}$ stress tensor has finite overlap with both the
ferro-orbital and the $d$- wave Pomeranchuk order parameters, and therefore criticality in either of these two channels is manifested in the Raman $B_{1g}$ response. Conversely, based on the Raman data, it is difficult to determine
whether the nematic criticality is ferro-orbital or $d$- wave Pomeranchuk type.

\subsection{Summary of the theoretical discussion}
\label{subsec:theory-summary}

Here we summarize the main points of the theory developed in the previous and in the current sections. In
Sec.~\ref{sec:nematic-theory} we developed a Drude-RPA theory to describe a charge nematic or $d$-wave
Pomeranchuk phase transition, starting from a phenomenological electron-electron interaction. In particular,
we showed that in purely electronic models the nematic susceptibility $\chi_n (\bq, \Om)$, that describes static and dynamic fluctuations of
the charge nematic operator $O_n(\bq)$, is analytic at zero momentum and frequency,
provided we take into account the effect of impurity induced elastic scattering of the electrons on their lifetimes.
The latter ensures that the single particle scattering rate does not vanish at zero frequency, which is a crucial ingredient
in order to establish the analyticity. We also noted that this
analyticity is eventually tied to the fact that the uniform charge nematic operator $O_n(\bq=0)$ is not a conserved
quantity. Next, in Sec.~\ref{sec:raman-theory} we discussed the characteristic signatures of the charge nematic instability
in the electronic Raman response. Within the Drude-RPA theory the effect of the criticality is symmetry-selective in the sense that
it is observed only in the Raman $B_{1g}$ channel, and not in the other Raman channels.  We showed that the $B_{1g}$ Raman response
$\chi^{\prime \prime}_{B_{1g}} (\Om)$ is
proportional to the imaginary part of the nematic susceptibility in its dynamical limit, i.e., to ${\rm Im} \chi_n (\bq = 0, \Om)$.
Next, using the analyticity discussed above, as well as the Kramers-Kronig relation, we argued that the frequency-integrated
Raman conductivity $\chi^{\prime \prime}_{B_{1g}} (\Om)/\Om$ is a measure of the nematic susceptibility at its static limit,
i.e., $\lim_{\bq \rightarrow 0} \chi_n (\bq, \Om = 0)$, which is the quantity whose divergence signals the second order charge nematic
transition. Furthermore, we showed that when the single particle lifetime is dominated by elastic scattering, the $B_{1g}$ Raman
conductivity has a Lorentzian lineshape whose width narrows as a function of temperature as the system approaches the nematic instability.
Finally, we pointed out that the electronic Raman response function naturally screens out the effect of the coupling of the electronic nematic
variable with the orthorhombic strain of the lattice. Consequently, the Raman response is a measure of the bare electronic nematicity that
is unaffected by the presence of the lattice. As such, it is an ideal tool for providing qualitative distinction between an electronically-driven
nematic instability from a lattice-driven one.

\section{Raman experiments}
\label{sec:Raman-exp}
\begin{figure}
\centering
\includegraphics[clip,width=0.79\linewidth]{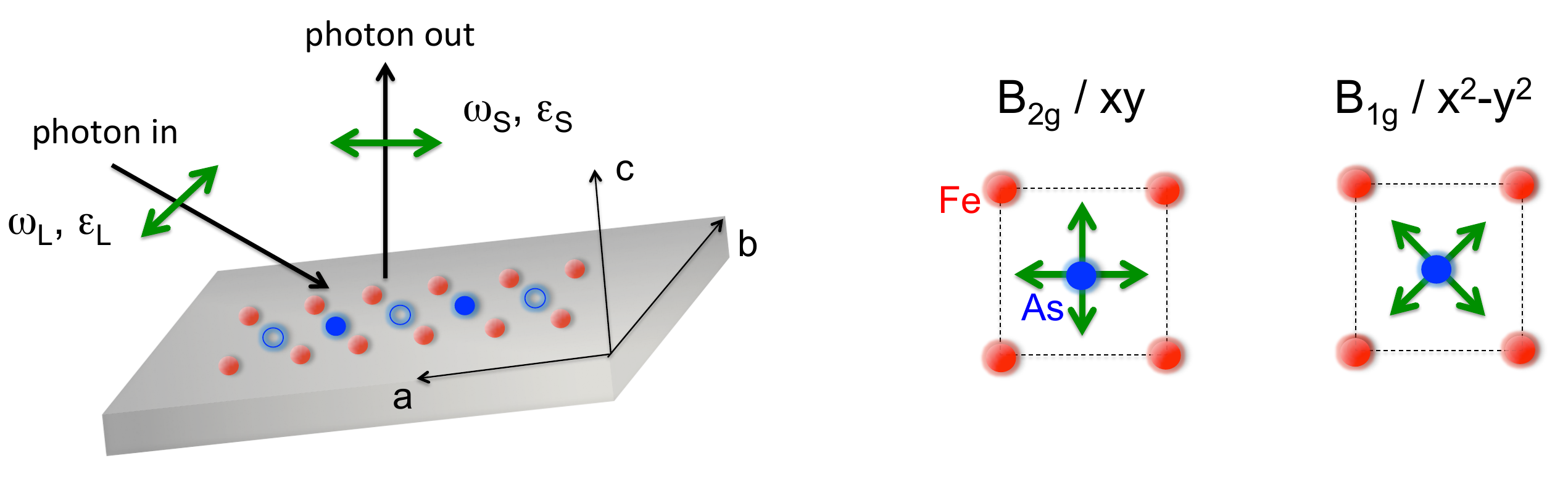}
\caption{Sketch of the scattering geometry in a Raman experiment. The photon polarization configuration corresponding to the $B_{1g}$
and $B_{2g}$ symmetries are depicted with respect to the FeAs plane. Note that the 1 Fe unit cell notation is used.}
\label{fig2}
\end{figure}

Raman scattering is a photon-in photon-out process in which an incident photon with energy $\omega_L$ and polarization $\epsilon_L$
is inelastically scattered by the medium into a photon of energy $\omega_S$ and polarization $\epsilon_S$ (Fig. \ref{fig2}). For a
Stokes process, $\omega_S<\omega_L$, an excitation with energy $\omega = \omega_L - \omega_S$ is created in the solid. $\omega$ is
usually referred to as the Raman shift and is traditionally given in units of cm$^{-1}$ (8.066~cm$^{-1}$=1~meV). A typical Raman
spectrum for a metal consists of sharp peaks due to Raman allowed optical phonons
superimposed on a continuum of electronic origin. The
Raman experiments described here were carried out using a diode-pumped solid state laser emitting at 532$\:\mathrm{nm}$ or a Ar-Kr mixed
gas laser with several lines in the visible spectral range. The inelastically scattered photon were analyzed using a triple
grating spectrometer equipped with a nitrogen cooled CCD camera. Special care was taken
in order to determine the laser induced heating. It was first estimated by comparing
the power and temperature dependencies of the phonon frequencies. This
estimate was then cross-checked by monitoring the onset of Rayleigh
scattering by orthorhombic structural domains across the structural
transition temperature as a function of laser power. For Co-Ba122 crystals both methods
yielded an estimated heating of 1$\:$K $\pm$ 0.2 per mW of incident
power. In order to extract the imaginary part of the Raman response
function, the raw spectra were corrected for the Bose factor using Eq. (\ref{eq:fluc-diss}) and for the
instrumental spectral response as well.

\par
In this review we will mostly discuss the spectra performed in the $B_{1g}$ symmetry because they probe the nematic degrees of freedom
relevant to the Fe SC.  The $x^{2}-y^{2}$ or $B_{1g}$ symmetry can be selected by choosing crossed incoming and outgoing photon
polarizations at 45 degrees with
respect to the Fe-Fe bonds. Here the notation $B_{1g}$ refers to
the one Fe unit cell whose axes are along the Fe-Fe bonds. For comparison we will also show spectra in $B_{2g}$ symmetry which can be
selected by choosing incoming
and outgoing photon polarizations along the Fe-Fe bonds. Switching between both symmetries was usually performed by rotating the crystal
by 45 degrees while keeping both the polarizer and the analyzer fixed. Note that in terms of the full lattice unit cell (or 2 Fe unit cell), which
has its axes at 45 degrees to the Fe-Fe bonds and is sometimes
used in the literature, the $B_{1g}$ ($B_{2g}$) symmetry discussed
here corresponds to the $B_{2g}$ ($B_{1g}$) symmetry.

\section{Experiments: Charge nematic susceptibility of electron doped 122 systems}
\label{sec:nematic-122}

\subsection{Electron doped Ba(Fe$_{1-x}$Co$_x$As)$_2$}
\label{subsec:nematic-CoBa122}

We first discuss Raman data in the normal state of electron doped Ba(Fe$_{1-x}$Co$_x$As)$_2$ (Co-Ba122) and the evolution of the extracted
static nematic susceptibility as a function of temperature and Co doping \cite{Gallais2013}. Being one of the most studied
Fe SC system, the salient features of the phase diagram as a function of Co doping are well-known \cite{Canfield2010}. There is a
quasi-simultaneous magneto-structural transition in the parent compound ($T_S\sim T_N$) \cite{Rotter2008,Huang2008} which splits under
Co doping with
$T_S> T_N$ \cite{Ni2008,Chu2009,Rullier-Albenque2009,Lester2009,Pratt2009,Kim2011}. The superconducting dome starts at $x\sim 0.02$
and extends up to at least $x$=0.15. All the Co contents $x$ mentioned in this review were determined using wavelength dispersing X-ray
spectroscopy (WDS).
\par
\begin{figure}
\centering
\includegraphics[clip,width=0.99\linewidth]{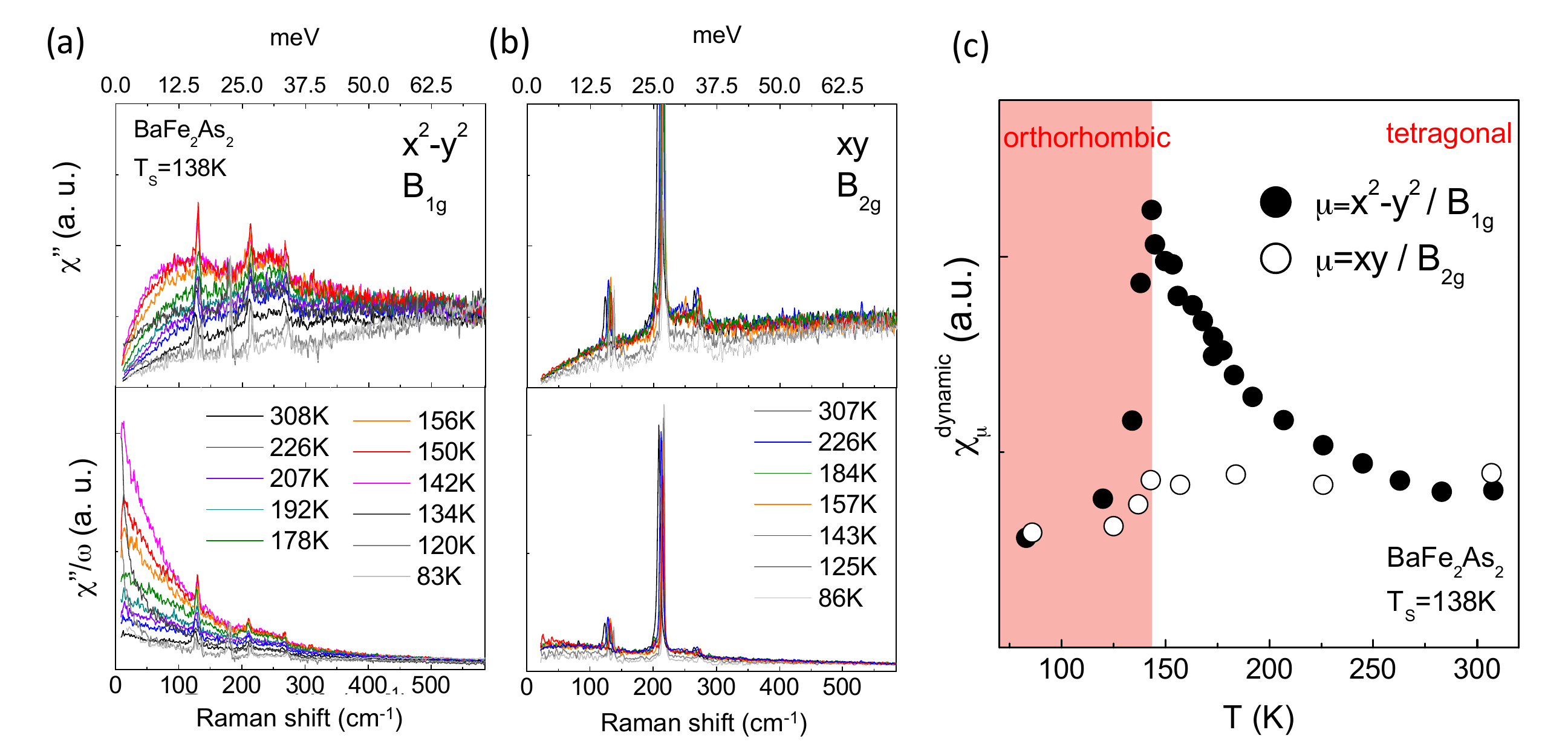}
\caption{Raman responses $\chi''$ and Raman conductivity $\chi''/\omega$ in the $B_{1g}$ (a) and $B_{2g}$ (b) symmetry of BaFe$_2$As$_2$.
The sharp lines superimposed on the electronic continuum are due to Raman active optical phonon excitations
(c) Temperature dependence of the extracted static nematic susceptibilities in the two symmetry channels using Kramers Kronig relation (see text).}
\label{fig3}
\end{figure}
\begin{figure}
\centering
\includegraphics[clip,width=0.99\linewidth]{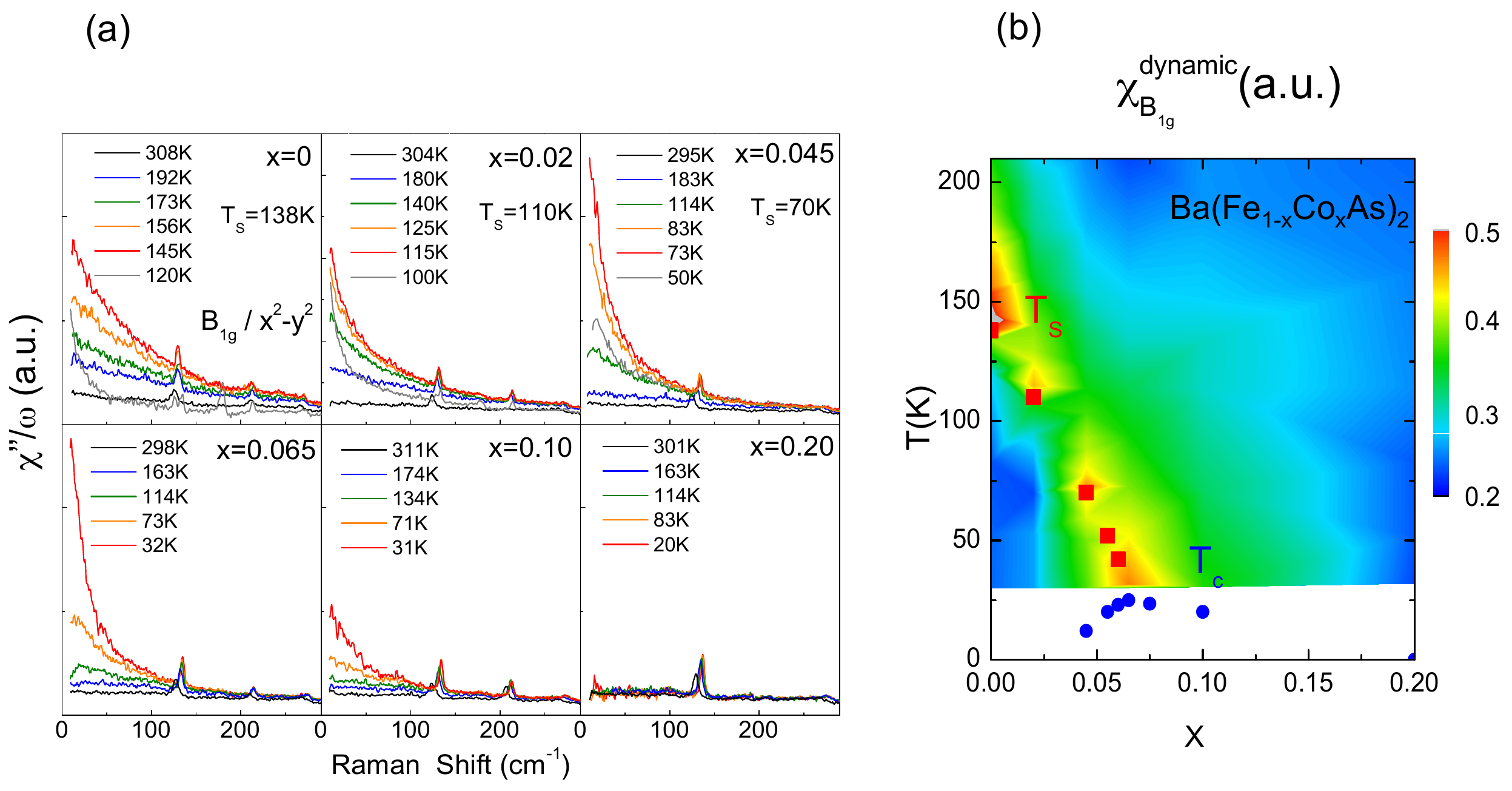}
\caption{(a) Temperature dependence of the Raman conductivity $\chi''/\omega$ in the $B_{1g}$ symmetry of Ba(Fe$_{1-x}$Co$_x$As)$_2$ (Co-Ba122)
for different Co doping concentrations $x$. (b) Color plot of the nematic susceptibility $\chi_{B_{1g}}^{dynamic}$ as function of
temperature and Co doping.}
\label{fig4}
\end{figure}

The low energy ($\omega <$ 600 ~cm$^{-1}$ or 75 ~meV) Raman responses of the parent compound BaFe$_2$As$_2$ in the $B_{1g}$ and $B_{2g}$
symmetries are shown as a function of temperature in Fig. \ref{fig3} (a) and (b). In the tetragonal phase the $B_{1g}$ response shows a
strong enhancement upon cooling towards $T_S$=138~K, before collapsing in the orthorhombic / spin density wave (SDW) state. By contrast
the $B_{2g}$ response is essentially independent of temperature in the tetragonal phase and only
shows a mild suppression below $T_S$. The observed symmetry dependence is in agreement with the presence of dynamical nematic fluctuations
having $x^2$-$y^2$ symmetry as discussed above. This interpretation is confirmed by the Co doping dependence of the spectra which shows a
systematic enhancement of the $B_{1g}$ response towards $T_S$ and the disappearance of any temperature dependence in the strongly electron
overdoped composition, far away from the orthorhombic instability ($x$=0.20, see Fig. \ref{fig4}).

\par
From the behavior of the Raman response at finite energy one can define the associated nematic susceptibility
$\chi_{\mu}^{\rm dynamic}$, where $\mu$ is the symmetry channel, using Kramers-Kronig relation (see Eq. \ref{eq:Kramers}). As emphasized
above, the susceptibility is in this case obtained in the dynamical limit. The physical quantity governing the nematic susceptibility in
this limit, the Raman conductivity $\chi''/\omega$, is shown as a function of symmetry for Ba122 in Fig. \ref{fig3}.
It is also shown as function of electron Co doping in Fig. \ref{fig4}. While the Raman conductivity is flat in $B_{2g}$ symmetry, in $B_{1g}$
symmetry it is dominated by a peak centered at zero energy over a wide range of Co doping in the tetragonal phase. For $x\leq$0.045 the peak
amplitude grows upon approaching $T_S$ and collapses quickly below. It is interesting to note that the enhancement of the peak amplitude is
also seen down to the superconducting phase, $T\sim T_c$, for $x$=0.065 and also, albeit more moderately, for $x$=0.10. At these two
compositions no $T_S$ is observed and the system remain tetragonal and paramagnetic down to
T=0~K \cite{Rullier-Albenque2009,Rullier-Albenque2010,Chauviere2009}. Data in the SC state will
be discussed in Sec. \ref{sec:nematic-SC}
\par
The Raman conductivity being independent of temperature for $\omega>$600~cm$^{-1}$ in the tetragonal phase, the quantity
$\chi_{\mu}^{dynamic}$ can be extracted in both symmetries by first extrapolating the data from the lowest energy
measured (9~ cm$^{-1}\sim$ 1~meV) down to 0~ cm$^{-1}$, and then integrating up to 600~cm$^{1}$. The symmetry dependence of this quantity
is shown in Fig. \ref{fig3}(c) for the parent compound Ba122. While the values of  $\chi_{\mu}^{dynamic}$ are very similar at high
temperature in $B_{1g}$ and $B_{2g}$ symmetries, the enhancement of the susceptibility upon cooling is only seen in $B_{1g}$ symmetry
clearly indicating an instability towards a charge nematic order with $B_{1g}$ or $x^2$-$y^2$ symmetry.
\par
The Co doping and temperature dependencies of $\chi_{B_{1g}}^{dynamic}$ are summarized in the color plot shown in Fig. \ref{fig4}(b).
The plot shows the clear correlation between the maximum of $\chi_{B_{1g}}^{dynamic}$ and $T_s$. It also highlights the persistence
of significant nematic fluctuations over a wide range of Co doping covering most of the superconducting dome.

\begin{figure}
\centering
\includegraphics[clip,width=0.99\linewidth]{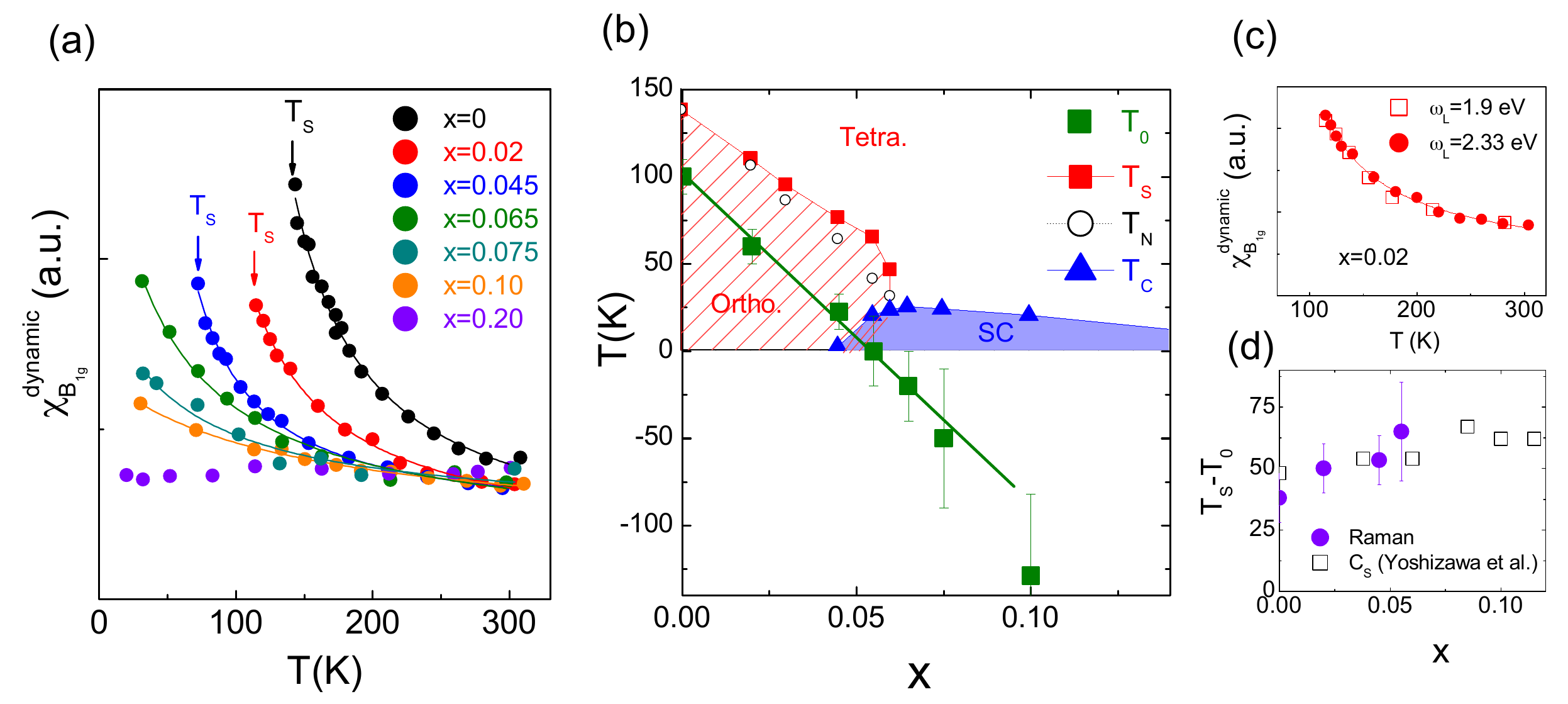}
\caption{(a) Curie-Weiss fits of the nematic susceptibility in the tetragonal phase of Co-Ba122. Structural transition temperatures
$T_S$ are marked by arrows for $x\leq$ 0.045. (b) Phase diagram showing the Co doping dependence of the bare electronic nematic
Curie-Weiss temperature $T_0$ (green squares). (c) Temperature dependence of the nematic
susceptibility deduced from measurements at two different incident photon energies fir $x$=0.02 ($T_S$=110~K). (d) Co doping dependence
of the difference between the measured structural transition temperature $T_s$ and $T_0$. The open squares correspond to the  same
quantity extracted from a Curie-Weiss analysis of the shear modulus $C_s$ by
Yoshizawa et al. \cite{Yoshizawa2012}}
\label{fig5}
\end{figure}

As shown in Fig. \ref{fig5}(a) for each Co composition the nematic susceptibility can be well fitted using a Curie-Weiss law
(see Eq.~\ref{eq:r0-2} in Sec.~\ref{subsec:instability} and Eq.~\ref{eq:R-dynamic-expt} in Sec. \ref{subsec:quasi-elastic}):
\begin{equation}
\chi_{B_{1g}}^{dynamic}=B+\frac{C}{T-T_0}
\label{CW}
\end{equation}
where $C=\frac{A_0}{\tilde{r_0}}$ is a temperature independent constant.

Here $B$ is a constant which describes the temperature and symmetry independent part of the susceptibility, i.e. the non-critical part of
the Raman response, and $T_0$ is the Curie-Weiss temperature corresponding to the lattice-free electronic nematic transition temperature
as defined in Sec.~\ref{subsec:instability}. The extracted values of $T_0$ follow qualitatively the doping dependence of $T_S$ and extrapolate
to zero slightly below optimal doping, at the critical doping $x$=$x_c \sim$ 0.055 (Fig. \ref{fig5}(b)). As displayed in Fig. \ref{fig5}
for $x$=0.02 the temperature dependence of the nematic susceptibility does not depend appreciably on the incident photon energy $\omega_L$
used for the Raman experiment. The insensitivity to $\omega_L$ suggests that the resonant terms in the Raman vertex, if present, do not
alter appreciably the temperature behavior of the $B_{1g}$ response above $T_S$, justifying the use of the effective mass  approximation
as done in Eq.~\ref{eq:gamma-k}
of \ref{subsec:B1g} and in Eq.~\ref{multi-vertex} of \ref{subsec:multi-orbital}.
\par
It is important to note that the $T_0$ values are systematically at least 40~K below the
actual thermodynamic structural transition temperature $T_S$.
This difference is a natural consequence of the absence of contribution of the lattice to the extracted susceptibility as discussed
in \ref{subsec:lattice}. $T_0$ represents the bare nematic transition temperature of the purely electronic system while in the presence
of the lattice, the actual transition temperature $T_S$ is moved to higher temperature due to the finite coupling
$\lambda_0$ between the electronic and lattice sub-systems (see Eq. \ref{eq:Tn}). In that case the observed divergence of the nematic
susceptibility is therefore cut-off by the structural transition which occurs at $T_S>T_0$. The difference between the two temperatures
is a measure of the charge-lattice coupling energy $\lambda_0^2/(c_S^0\tilde{r_0}g_0)$. This quantity, which is accessible both from
Raman and from elastic constant measurements, is shown as function of Co doping in Fig. \ref{fig5}(d).

\subsection{Comparison with elastoresistivity and elastic measurements}
\label{subsec:exp-comp}
The observed Curie-Weiss like enhancement of the charge nematic susceptibility is qualitatively consistent with both elastoresistivity and
elastic modulus measurements performed on Co-Ba122 \cite{Chu2012,Goto2011,Yoshizawa2012,Bohmer2014}. The
presence of orbital fluctuations above $T_S$ was also inferred from point-contact spectroscopy measurements in the same system \cite{Arham2012}.
In the case of elastoresistivity measurements the purely electronic nematic susceptibility could be obtained by strain dependent measurements
of the resistivity anisotropy in the tetragonal phase \cite{Chu2012}. The extracted divergence was taken as evidence for a electronic driven
structural transition. Since the nematic susceptibility extracted from Raman measurements
is free from lattice effects, they also confirm the present of diverging electronic nematic degrees of freedom in the tetragonal phase of
undoped and electron doped Ba122. We note however that the Curie-Weiss temperatures extracted from transport measurements are significantly
higher than the ones obtained from Raman measurements on samples with similar Co doping. The discrepancy could be due to different couplings
to electronic nematic degrees of freedom. Indeed, while Raman scattering couples to charge nematic degrees of freedom, the nematic component
of the elastoresistivity tensor is a more complex quantity.
On the one hand it could be a measure of Drude weight anisotropy arising from the sensitivity of the electronic structure to generate
anisotropy in the presence of external strain~\cite{Valenzuela2010}. On the other hand, elastoresistivity is also sensitive to anisotropy
in transport lifetimes of the carriers that can arise from scattering with the spin fluctuations or the
impurities~\cite{Fernandes2011,Blomberg2013, Ishida2013,Kuo2014,Breitkreiz2014,Gastiasoro2014,Wang2014}.
Deviations from Curie-Weiss behavior have also been recently reported in several Fe SC near optimal doping by elastoresistivity
measurements \cite{Kuo2015}. They have been interpreted as due to random field disorder that is a relevant perturbation near a quantum
critical nematic transition.

\par
Elastic modulus measurements (discussed in a separate contribution in this issue \cite{Bohmer-CRAS}) can also be fitted with a
Curie-Weiss like temperature dependencies.  While Young's modulus measurements from three point bending technique are now available
for different Fe SC systems and dopings \cite{Bohmer2014,Bohmer2015}, direct measurements of the shear modulus $C_S$ from
ultrasound velocity measurements  are only available for Co-Ba122 \cite{Fernandes2010,Goto2011,Yoshizawa2012}. In both types of measurements
a Curie-Weiss like softening is observed in Co-Ba122. The extracted Curie-Weiss temperatures are very close to $T_S$, which is consistent
with a second order structural phase transition. We note that recent neutron scattering measurements in parent 122 compounds also observe a clear softening of the associated transverse acoustic phonon at low wavevectors \cite{Parshall2015}.
Assuming a simple Landau type approach of the coupling between shear modulus and electronic nematic degrees of freedom identical to the one presented in \ref{subsec:lattice}, the bare electronic
nematic transition temperature could also be extracted from these elastic measurements using the relation $C_S=C_S^0(\frac{T-T^{CW}_S}{T-T_0})$, where $T_S^{CW}\sim T_S$ and $T_0$ is the purely electronic nematic transition temperature. The $T_0$ values obtained from shear modulus
measurements in Co-Ba122 agree remarkably well with the ones extracted from Raman
measurements \cite{Yoshizawa2012, Gallais2013} \footnote{We note that the $T_0$ extracted from Young's modulus measurements are
somewhat higher for Co-Ba122 \cite{Bohmer2014}. This might be due to the contribution of other non-critical components of the elastic
tensor to the Young's modulus.}. For both Raman and shear modulus measurements $T_S$-$T_0$ increases mildly with Co doping
(see Fig. \ref{fig5}(d)), indicating a possible increase of the electron-lattice coupling energy scale $\frac{\lambda_0}{C_s^0\tilde{r}_0g_0}$
(see Eq. \ref{eq:Tn}) upon electron doping. We will come back in \ref{subsec:Raman-shear}
for a more quantitative comparison between elastic and Raman measurements.

\subsection{Impact of disorder: Sr(Fe$_{1-x}$Co$_x$As)$_2$}
\label{subsec:disorder}
As already stated above, mechanisms of anisotropic scattering are a possible source of the transport
anisotropies~\cite{Fernandes2011,Breitkreiz2014,Gastiasoro2014,Wang2014},
associated with lifetime effects, that have been reported in both
the orthorhombic and the tetragonal phases under applied strains or stress. Indeed transport measurements performed on annealed crystals
appear to show a much reduced anisotropy in the orthorhombic phase, hinting at a key role of scattering mechanisms in the observed
nematicity \cite{Ishida2013}. STM measurements have further shown the nucleation of nematic nano-domains around defect sites in the
orthorhombic, and possibly even in the tetragonal phase \cite{Allan2013,Rosenthal2014}. These anisotropic impurity states were postulated
to be responsible for the observed transport anisotropies because they are expected to act as strongly anisotropic
scatterers \cite{Gastiasoro2014,Gastiasoro2014b,Sugimoto2014}. Recent NMR measurements also suggest the presence of short range but static
nematic order above $T_S$. This short range order could be due to pinning by impurities or to the presence micro-strains \cite{Iye2015} and
could possibly explain the onset of anisotropy observed at $T>T_S$ in magnetic torque measurements \cite{Kasahara2012}.
\par
\begin{figure}
\centering
\includegraphics[clip,width=0.99\linewidth]{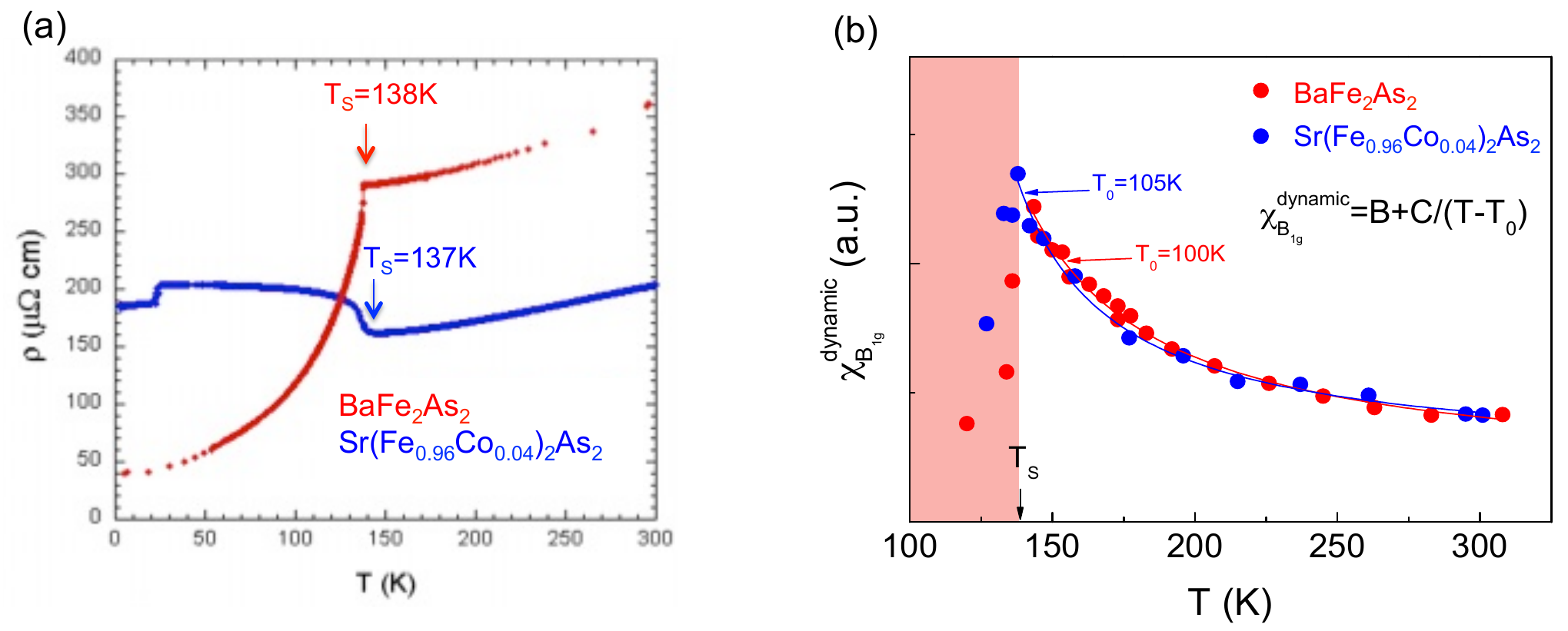}
\caption{(a) Temperature dependence of the resistivity of Ba122 (red, RRR$\sim$ 9) and Co-Sr122 (x=0.04, blue, RRR$\sim$1). Unpublished
data courtesy of F. Rullier-Albenque. (b) Corresponding nematic susceptibilities extracted from Raman measurement in the $B_{1g}$
symmetry \cite{Yang-SCES}. The lines are Curie-Weiss fits of the data points.}
\label{fig6}
\end{figure}

The above measurements have raised the question of the intrinsic nature of nematicity in Fe SC. A direct comparison between the Raman
measurements in Ba122 ($T_S$=138K) and Co-Sr122 (x=0.04, $T_S$=137K) allows a direct assessment of the possible role of disorder in the
emergence of nematic fluctuations above $T_S$ \cite{Yang-SCES}. Indeed despite having essentially identical $T_S$,
resistivity measurements shown in Fig. \ref{fig6}(a) indicate a much higher residual resistivity ratio (RRR) in Co-Sr122 likely caused by
the insertion of Co in the FeAs plane. Despite an order of magnitude difference in RRR, the extracted nematic susceptibilities shows extremely
similar temperature dependencies
in both systems (see Fig \ref{fig6} (b)). In particular the extracted Curie-Weiss temperatures $T_0$ agree within $\pm$5~K, clearly
demonstrating a relative insensibility of the nematic fluctuations and their associated diverging susceptibility to disorder. The Raman
measurements are consistent with recent elasto-resistivity measurements which show that while the magnitude of the observed transport
anisotropies might be disorder dependent, the diverging behavior of the extracted susceptibility in the tetragonal
state is not \cite{Kuo2014}. This conclusion is supported by a recent strain-dependent optical conductivity study which suggests that the strain induced transport anisotropy observed above $T_S$ is not due to anisotropic scattering rate, but rather to anisotropic Drude weight \cite{Mirri2015}. It is likely that disorder helps revealing an underlying intrinsic nematicity in transport and local probe
measurements. The magnitude of the observed anisotropies is thus not necessarily the right quantity to assess the intrinsic nematicity
of a given Fe SC system.

\section{Dynamical aspects of the charge nematic response}
\label{sec:nematic-dynamics}
\subsection{Quasi-elastic peak}
\label{subsec:QEP}
\begin{figure}
\centering
\includegraphics[clip,width=0.99\linewidth]{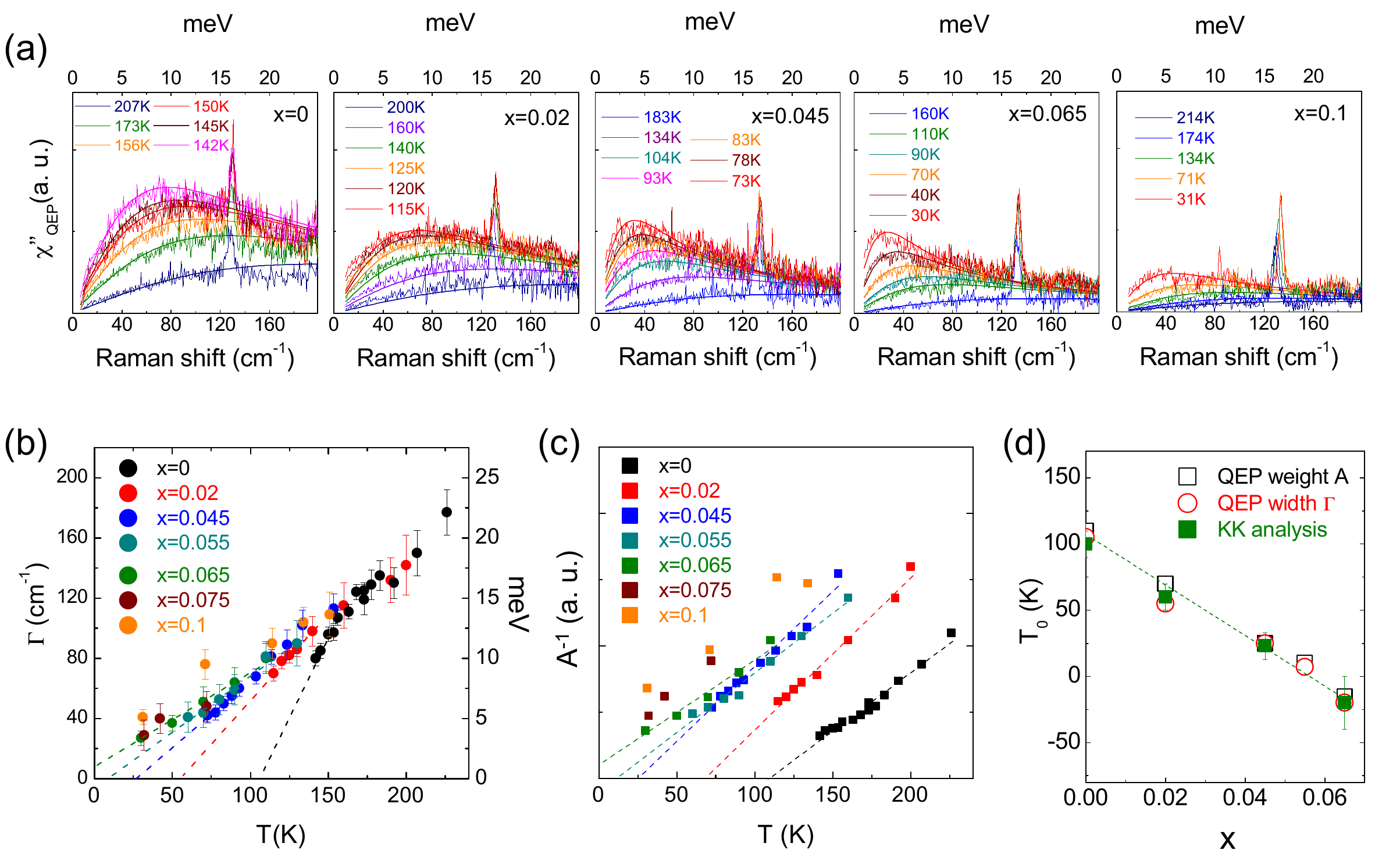}
\caption{(a) Evolution of the quasi-elastic peak contribution (QEP) in the $B_{1g}$ symmetry as a function of temperature and Co doping.
The contribution was extracted by subtracting the raw $B_{1g}$ response by the one in $B_{2g}$ symmetry for each doping. Fits using a
damped Lorentzian are shown for each spectra. (b) Temperature evolution of width $\Gamma$ of the quasi-elastic peak. (c) Temperature
dependence of the inverse of the area $A^{-1}$ of the QEP. (d) Co doping evolution of the mean-field nematic transition $T_0$ extracted
by three different methods.}
\label{fig78}
\end{figure}

We now go beyond the static properties of the nematic susceptibility
discussed in the previous section, and analyze the frequency dependence of the
nematic fluctuation spectrum revealed by the Raman measurements in $B_{1g}$ symmetry. The temperature and symmetry dependencies of
the Raman response in Ba122 (Fig. \ref{fig3}) suggest the presence of two contributions. The first one, broad in energy and weakly
temperature dependent, is seen in both symmetries and dominates the Raman responses at high temperature. It can be assigned to intra
and/or interband quasiparticles excitations which do not become critical upon cooling. This contribution shows only a weak suppression
in the orthorhombic phase, which is linked with the simultaneous opening of the SDW gap. The second contribution, only present in
$B_{1g}$ symmetry, is strongly temperature dependent and is responsible for the strong enhancement of the static nematic susceptibility
as discussed above. This contribution, obtained by subtracting the first contribution, is shown as a function of Co doping in
Fig. \ref{fig78}(a).
The observed dynamical nematic fluctuations are quasi-elastic in nature and can be well reproduced by a damped Lorentzian lineshape
of width $\Gamma$ and amplitude $A$ in agreement with the form obtained in the theory part of this review
(Eq. \ref{eq:R-low-Im}):

\begin{equation}
\chi''_{QEP}=A\frac{\omega \Gamma}{\omega^2+\Gamma^2}.
\label{QEP}
\end{equation}

The temperature dependence of the width of the quasi-elastic peak (QEP) $\Gamma$ is shown in Fig. \ref{fig78}(b). We find that
$\Gamma$ softens considerably upon approaching $T_S$ for $x \leq$ 0.045. This is because $\Gamma$ is approximately the single
particle lifetime renormalized by $(a/\xi_n)^2$, where
$a$ is the unit cell length and $\xi_n$ is the nematic correlation length, and the latter starts to increase
as the system approaches the nematic instability with lowering temperature. Assuming a Curie-Weiss $T$ dependence of the nematic
susceptibility, and sufficiently close to $T_0$, the temperature dependence
of $\Gamma$ is expected to vanish linearly with a zero intercept at $T_0$, the bare electronic nematic transition
temperature (see Eq. \ref{eq:Gamma} in \ref{subsec:quasi-elastic}):
\begin{equation}
\Gamma= \frac{r_0}{(r_0+c_3)\tau} \propto (T - T_0).
\end{equation}
For all Co compositions $\Gamma$ is found to decrease linearly between $T_S$ and up to at least 40K above $T_S$. As in the case of
the static susceptibility, the softening of $\Gamma$ is not complete because of lattice effects which move the transition temperature
to a higher temperature $T_S$.
\par
 The above analysis also allows us to extract the temperature dependence of the QEP area $A$, which is directly proportional to
 the diverging part of the nematic susceptibility i.e. without the background non-singular contribution
 $B$ in Eq. \ref{CW}.
 \begin{equation}
 \chi_{B_{1g}}^{dynamic, QEP}=\frac{2}{\pi}\int^{\infty}_0 \frac{d\omega}{\omega} \chi''_{QEP}(\omega) = A
 \end{equation}
As shown in Fig. \ref{fig78}(c), the temperature dependence of $A^{-1}$ is fairly linear, as expected for a Curie-Weiss
behavior, with $A^{-1}\sim (T - T_0)$. Extrapolation of the measured temperature dependencies
of $\Gamma$ and $A^{-1}$ in the tetragonal phase down to zero temperature provide a measure of the scale $T_0$,
which is in good agreement with the values obtained from the Curie-Weiss fits of the static susceptibility
described in the previous section (see Fig. \ref{fig78}(d)).

\subsection{Adequacy of RPA theory of the nematic transition}
\label{subsec:RPA}
\begin{figure}
\centering
\includegraphics[clip,width=0.99\linewidth]{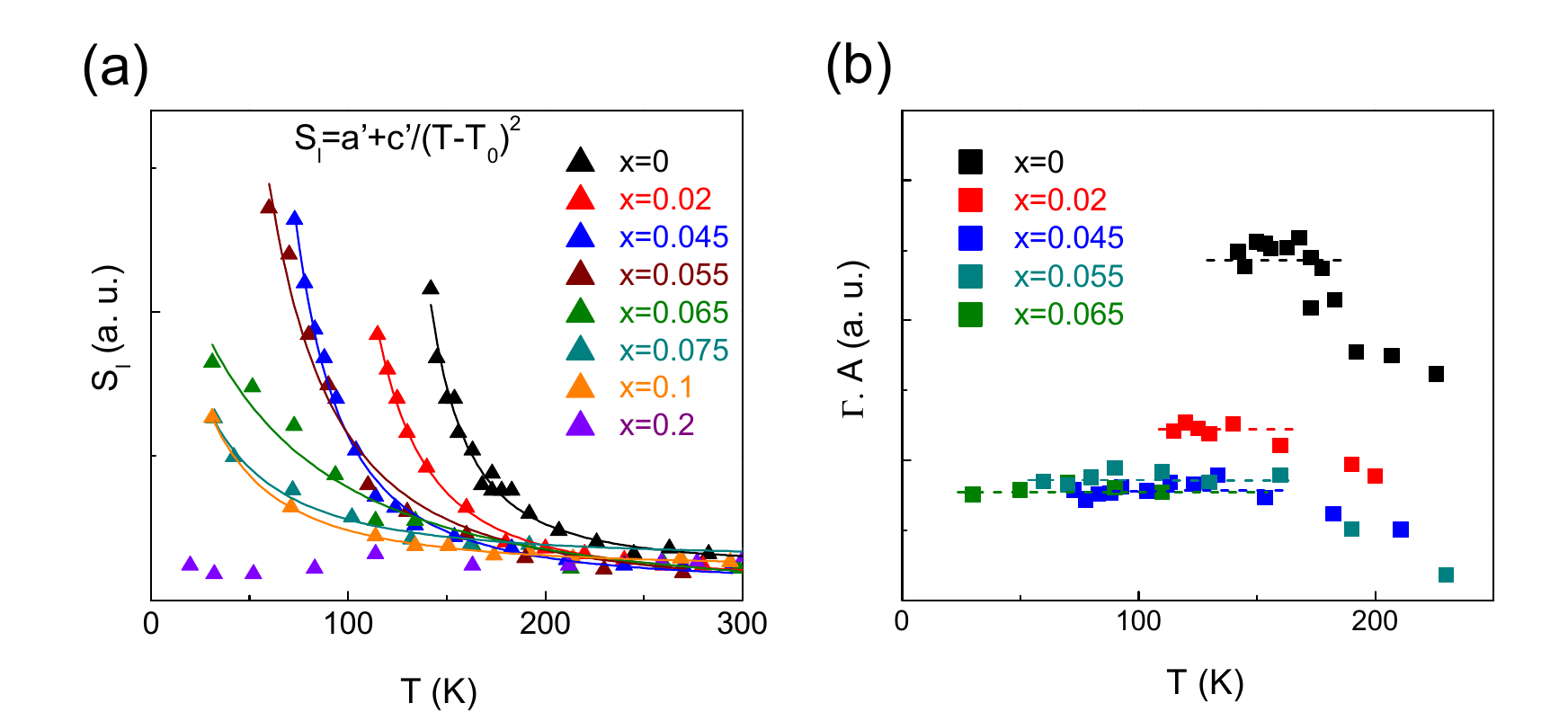}
\caption{(a) Temperature dependence of the slope of the Raman response at low energy $S_l$. The solid lines are fits
using $S_l= a'+\frac{c'}{(T-T_0)^2}$. (b) Temperature dependence of the product $\Gamma \dot A$ where $\Gamma$ is the
effective one-particle scattering rate and $A$ the diverging part of the nematic susceptibility (see text) }
\label{fig9}
\end{figure}

Both the frequency and the temperature dependencies of the nematic susceptibility are in broad agreement with the expectations from
the simple RPA description of a charge nematic transition outlined in the theory section of this review. The fact that the diverging
nematic susceptibility is interrupted by the structural transition is a direct consequence of the lack of coupling to the lattice of
a nematic susceptibility extracted from Raman in the dynamic limit. Both the Curie-Weiss behavior of the susceptibility and the linear
decrease with $T$ of the effective one particle scattering rate $\Gamma$ are also in agreement with a simple mean-field description
of a charge nematic instability described in \ref{subsec:quasi-elastic}.
The RPA type picture can be further tested by looking at the low energy slope of the $B_{1g}$ Raman response, $S_l$, which is
predicted to vary as $(T-T_0)^{-2}$ (see Eq. \ref{eq:S_l}). As illustrated in Fig. \ref{fig9}(b) the quantity $S_l$ follows nicely
the expected behavior for all Co dopings, even when forcing $T_0$ to the value previously obtained from the Curie-Weiss fits of the
susceptibility.
\par
A further test of the RPA picture can be made by looking at the temperature dependence of the product $\Gamma . A$. Identifying
Eq. \ref{eq:R-low-Im} with Eq. \ref{QEP}, this product should become almost  temperature independent upon approaching $T_0$:
$\Gamma A\sim \frac{A_0}{\tau}\sim \frac{1}{g_0\tau}$. This is indeed what is found over a wide range of Co composition as shown
in Fig. \ref{fig9}(b). The decrease in $\frac{1}{g_0\tau}$ from $x=0$ to $x$=0.045 can come from different sources. It could be
linked to the decrease of the bare quasiparticle scattering rate $\frac{1}{\tau}$ as suggested by transport measurements in Co-Ba122
\cite{Chu2009,Rullier-Albenque2009}, but it could also be due to an increase of the coupling constant $g_0$ upon Co-doping.
Overall there is, thus, a remarkable consistency between the behavior of the nematic $B_{1g}$ Raman response in Co-Ba122 and the
theoretical expectations from a simple RPA theory of the nematic transition.

\subsection{A consistent picture with shear modulus measurements}
\label{subsec:Raman-shear}
\begin{figure}
\centering
\includegraphics[clip,width=0.99\linewidth]{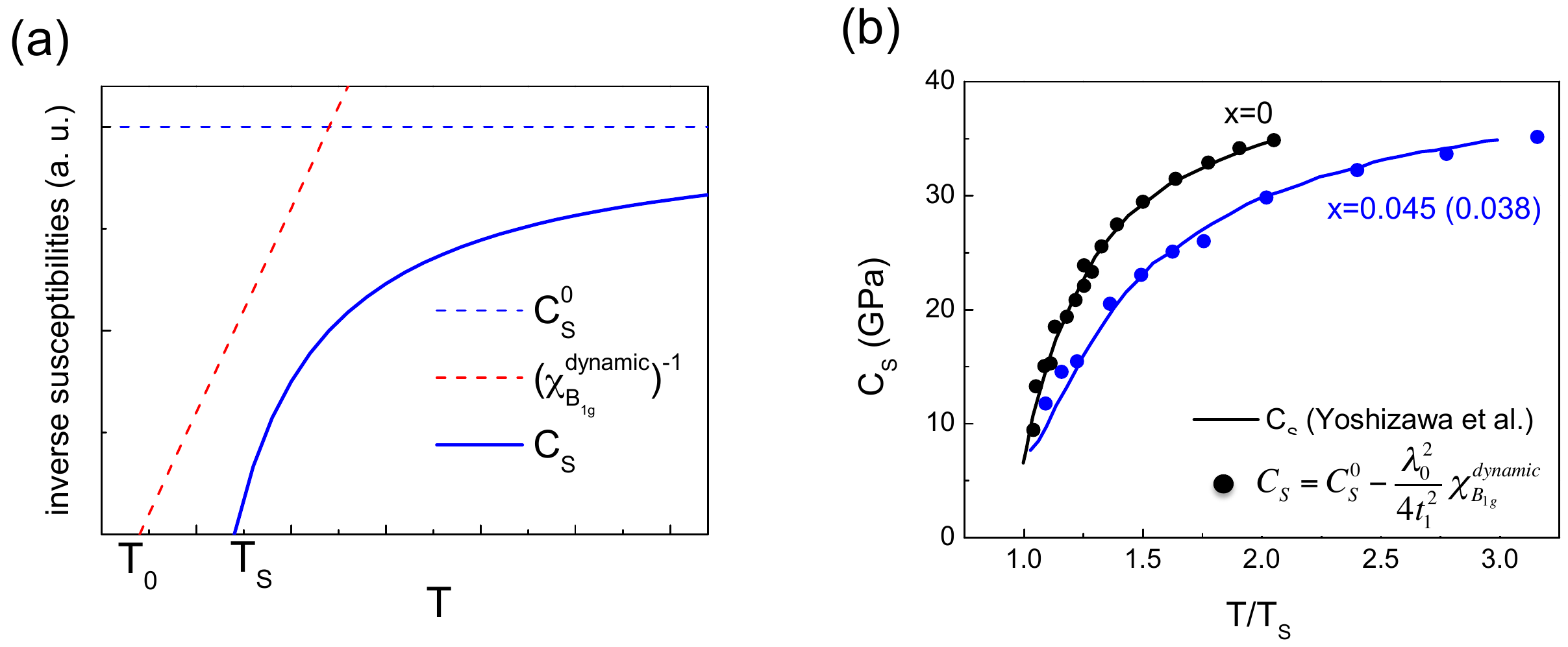}
\caption{(a) Mean-field model of the linear coupling between the inverse lattice (or shear modulus) and electronic nematic susceptibilities.
The bare shear modulus $C_S^0$, initially independent of temperature, is softened by the linear coupling to a diverging nematic electronic
susceptibility $\chi^{dynamic}_{B_{1g}}$ causing a phase transition at $T_S>T_0$ (b) Comparison between the experimental temperature
dependence of the shear modulus $C_S$ \cite{Yoshizawa2012} for Co-Ba122 ($x$=0 and $x$=0.038) and the theoretical temperature dependence
using the charge nematic susceptibility deduced
from Raman measurements ($x$=0 and $x$=0.045, see text for details). To account for the slight mismatch in the structural transition
temperatures $T_S$ of the crystal used in the two experiments, the temperature axis was set in units of $T_S$.}
\label{fig10}
\end{figure}

The Raman measurements indicate the presence of genuine electronic nematic fluctuations in the tetragonal phase of Co-Ba122. These
fluctuations are expected to soften the lattice and trigger the structural
transition \cite{Fernandes2010,Cano2010}. The availability of both Raman and  shear modulus data in Co-Ba122 allows us to draw a more
quantitative picture linking electronic and lattice degrees of freedom. Indeed by symmetry the charge nematic order parameter $O_n$
probed by Raman and the orthorhombic lattice distortion $\epsilon_O$ are linearly coupled (Eq. \ref{eq:el-lattice}).
Remembering that Raman probes the bare charge nematic susceptibility, and using Eqs.~(\ref{eq:R-B1g-2}) and (\ref{eq:Cs}),
we have a simple relationship between
$\chi_{B_{1g}}^{dynamic}$ and $C_S$:
\begin{equation}
C_S=C_S^0-\frac{\lambda_0^2}{4 t_1^2}\chi_{B_{1g}}^{dynamic}
\label{Raman-shear}
\end{equation}
where $\lambda_0$ is the charge lattice coupling constant introduced in \ref{subsec:lattice} and $t_1$ is the nearest-neighbor
hopping parameter introduced after Eq.~(\ref{eq:gamma-k}). Assuming a weakly
temperature dependent $C_S^0$ the renormalized shear modulus can become soft via the linear coupling to the charge nematic susceptibility
$\chi_{B_{1g}}^{dynamic}$ causing a second order structural phase transition at $T_S$ defined as $C_S(T=T_S)=0$ (see Fig. \ref{fig10}(a)).
Fig. \ref{fig10}(b) shows a comparison between the shear modulus data of Yoshizawa et al. \cite{Yoshizawa2012} and the theoretical shear
modulus expected from the Raman data using equation \ref{Raman-shear} using $\lambda_0$ as the only free parameter. The quantitative
agreement between shear modulus data and the theoretical expectation for both undoped and Co doped Ba122 demonstrates that the charge
nematic fluctuations observed by Raman scattering can fully account for the structural softening observed in Co-Ba122. It also provides
a further evidence that the structural distortion observed in Fe SC is electronic driven. We stress that such quantitative agreement
cannot be reached if we assume that Raman scattering couples to the fully dressed nematic susceptibility i.e. including lattice effects.
\emph{The realization that Raman probes the bare nematic susceptibility is therefore crucial in reaching a consistent picture.}
\par

One question left open in our discussion until now is the role of spin degrees of freedom in the nematicity observed in Fe SC. Anisotropic
spin fluctuations associated with the stripe AF instability at or below $T_S$ of many Fe SC have been invoked early on as a main driver
for nematicity \cite{Xu2008,Fang2008}. Naively one might conclude from the consistency between shear modulus and Raman measurements that
only charge / orbital degrees of freedom as probed by Raman scattering are relevant in driving the structural transition. This would
however be incorrect, since it is very plausible that the diverging charge nematic susceptibility
is itself enhanced or even driven by spin-nematic fluctuations via spin-charge
coupling \cite{Onari2012,Liang2013,Paul2014,Yamase2015,Khodas2015, Karahasanovic2015,Kretzschmar2015}. In fact it has been argued
that a similar consistency as in Fig. \ref{fig9}(c) can be reached by comparing NMR measurements of spin fluctuations at
$Q_{SDW}$=(0,$\pi$)/($\pi$,0) and shear modulus data \cite{Fernandes2013}. Disentangling between the electronic degrees of freedom behind
nematicity thus remains
a formidable task in 122 systems \cite{Fernandes-Review-2014}. Other systems like FeSe with no magnetic
transition \cite{McQueen2009,Imai2009,Nakayama2014,Baek2015,Bohmer2015, Watson2015, Massat2015b} , or hole doped Ba122 with a still mysterious
$C_4$ magnetic phase \cite{Avci2014,Khalyavin2014,Bohmer2015b,Allred2015} may prove
fruitful playgrounds to answer this question. It is also possible that, like the structure of the SC gap, the respective weights of various
electronic degrees of freedom behind nematicity are not universal and depend on the Fe SC systems considered.

\subsection{Digression: Raman fingerprints of nematicity in cuprates}
The presence of nematic correlations as possible precursor of uni-axial modulated charge / spin ordered states is also intensively
discussed in the context of cuprates \cite{Tranquada1995, Ando2002, Daou2010, Lawler2010, Wu2011,Ghiringhelli2012,Chang2012}. It is
therefore interesting to discuss here earlier Raman data on La$_{1-x}$Sr$_x$CuO$_4$ (La-214) and YBa$_2$Cu$_3$O$_{6+\delta}$ (Y-123) which
are somewhat reminiscent of the results discussed here, albeit with some key
differences \cite{Tassini2005,Caprara2005,Tassini2008,Yamase2011,Caprara2015}. In underdoped La-214 a Raman QEP was observed to grow upon
cooling over a relatively broad doping range \cite{Tassini2005}. The symmetry of the QEP was found to switch from B$_{2g}$ to B$_{1g}$
channels at the critical doping $p$=0.05, indicating a 45 degrees rotation of the nematic fluctuations: along the Cu-O bond at moderate
doping and at 45 degrees of the Cu-O at low doping. This is in agreement with neutron scattering studies where stripe-like spin modulations
were also found to rotate upon increasing doping \cite{Wakimoto1999}. In La-214 there appears thus to be an intimate link between nematic
($q$=0) fluctuations observed by Raman and finite $q$ stripe-like charge / spin fluctuations \cite{Caprara2005,Caprara2011}. This
phenomenology is similar to the spin-nematic scenario for the structural transition in Fe SC discussed above.
In contrast with underdoped Co-Ba122 however, the growth of the QEP in La-214 and its softening saturate at low temperature and
therefore no true static long range order sets in \cite{Tassini2005}. Besides the growth of the QEP in La-214 is accompanied by the
opening of a pseudogap at intermediate energies upon lowering temperature whose relationship with nematicity is still unclear.
\par
In other cuprates like Y-123, Bi$_2$Sr$_2$CaCu$_2$O$_{8+\delta}$  and HgBa$_2$Cu0$_{4+\delta}$  evidence for nematic fluctuations in
the Raman spectra are more elusive as no clear QEP is observed for moderately underdoped systems in any
symmetry \cite{Opel2000,Gallais2005,Tassini2008}. Notably in moderately underdoped Y-123 where CDW fluctuations are clearly observed
in X-ray measurements and where transport anisotropies have been reported, no QEP is observed in the nematic $B_{1g}$ and B$_{2g}$
symmetries \cite{Opel2000,Tassini2008}. It has been recently suggested that the opening of the pseudogap may suppress the nematic QEP
in $B_{1g}$ symmetry in moderately underdoped cuprates \cite{Caprara2015}. We also note that in Y-123, whether the incipient order is
nematic or bi-axial is still
controversial \cite{Leboeuf2013, Sebastian2014, Tabis2014,Blanco-Canosa2014, Hucker2014,Fujita2014,Comin2015,Doiron2015,Forgan2015}.
By contrast a clear QEP emerges in $B_{2g}$ symmetry for strongly underdoped composition ($p < 0.05$) \cite{Tassini2008}. In this regime
however neutron scattering data indicate the presence of a uni-axial magnetic correlations along the Cu-O bond rather the Cu-O diagonal
as suggested by the $B_{2g}$ symmetry of the Raman QEP \cite{Haug2010}. Here again the connection between Raman results and other
probe appears problematic. The role of nematicity in cuprates and its link to the pseudogap is therefore largely unsettled as many
competing and fluctuating spin and/or charge ordered phases are in close proximity.

\section{Nematicity and superconductivity}
\label{sec:nematic-SC}

As of date, the study of nematicity in the Fe SC has concentrated mostly on the role of the various electronic nematic degrees of freedom
in driving the structural transition in the normal metallic state. Relatively less attention has been given to study the nature of these
fluctuations, if at all they exist, in the superconducting phase. In the following we show that, in a fully-gapped superconductor, soft
nematic fluctuations lead to a new collective mode whose signature is a finite frequency resonance peak in the electronic Raman response.
This behavior is qualitatively different from the quasi-elastic peak that develops when the nematic transition is approached in the normal state.

A second motivation to study nematic fluctuations in the superconducting phase comes from the possibility that
such fluctuations might affect the pairing mechanism and the superconducting transition.
Theoretically, it has been argued that nematic fluctuations can play a role in the interplay between $s$ and $d$-wave pairing states which are
likely nearly degenerate in several Fe SC
systems~\ \cite{Kuroki2009,Wang2009,Graser2009,Graser2010,Thomale2011,Maiti2011b, Maiti2011,Fernandes2013b}. Concerning the effect
on pairing, recent theoretical works~\cite{Maier2014,Lederer2015} suggest that nematic fluctuations enhance superconductivity irrespective of
the gap symmetry, at least for weak coupling
superconductors. Therefore, it is possible that nematic fluctuations provide a complementary contribution to the pairing interaction on top of
the $(\pi$,0) magnetic fluctuations that promote $s^{\pm}$ pairing symmetry.
\par
On the experimental side, both X-ray measurements of the orthorhombicity and elastic modulus measurements from ultrasound velocity indicate
competition between nematic and superconducting order parameters, which is in broad agreement with the expectations from simple
Ginzburg-Landau-type theoretical arguments~\cite{Nandi2010,Fernandes2010,Goto2011,Moon2010}. However, these studies did not address the
question whether the nematic/orthorhombic transition stays second order (or weakly first order) in the superconducting
phase~\cite{Fernandes2013c}, and if yes, how the properties of the accompanying nematic fluctuations differ from those in the normal phase.
In fact, experimental evidence for the existence of nematic fluctuations in the superconducting phase was lacking until recently.
Consequently, it is notable that a recent study has identified resonance peaks in
the electronic Raman response of Co-doped Ba122 and Na111 as evidence of the existence of nematic fluctuations in the superconducting
phase~\cite{Gallais2015}.

With the above motivations, in this section we briefly discuss the theory of nematic fluctuations in a SC phase, and we compare it
with the experimental findings from Raman response measurements in the SC state of electron doped FeAs. We finish by comparing the
properties of the nematic resonance mode with those of other collective modes that have been postulated to exist in the context of
superconductivity in the Fe SC.

\subsection{Theory: Nematic resonance near a nematic quantum critical point}
\label{subsec:nematic-resonance-theory}

In the superconducting phase the nematic fluctuations can be modeled by the same RPA
approach as the one used to describe the QEP in the normal state data.
As in the normal state, we assume that the nematicity is driven by electronic interaction of the form given by
Eq.~(\ref{eq:hamI}), with the interaction constant $g_0$ replaced by $g$, its value in the superconducting phase.
Within RPA the interacting nematic susceptibility probed in $B_{1g}$ symmetry in the superconducting state can be expressed in term of
the bare superconducting response and $g$. It is given by (see Eq. \ref{eq:chi-n} and
Eq. \ref{eq:R-B1g-1})
\begin{equation}
\chi_{B_{1g}}(\omega)=\frac{\Pi_{B_{1g}}(\omega)}{1-g \Pi_{B_{1g}}(\omega)}.
\label{eq:nematic-res}
\end{equation}
Mathematically it is clear that, since in the superconducting phase the Bogoliubov excitations are gapped, the
structure of $\Pi_{B_{1g}}$ is quite different from that of a normal metal.
For a fully gapped clean superconductor there are no fermionic excitations below  2$\Delta$. Consequently,
the imaginary part of the bare nematic response $\Pi_{B_{1g}}^{\prime \prime}$ is zero below  2$\Delta$,
and it is dominated by a pair-breaking
peak at 2$\Delta$. It is simple to check that, by Kramers-Kronig relation, this implies that
the real part $\Pi_{B_{1g}}^{\prime}$ diverges logarithmically upon approaching 2$\Delta$ from below. Therefore for positive nematic
coupling $g$ the RPA response develops a resonance \textit{below $2\Delta$} at an energy $\Omega_r$ defined as
\begin{equation}
1-g \Pi_{B_{1g}}^{\prime}(\Omega_r)=0.
\end{equation}
This condition will always be met for a fully gapped superconductor in the $C_4$-symmetric phase
near the nematic instability. Thus, the presence of a nematic resonance in the Raman spectra is a generic property of the superconducting
excitation spectrum near a nematic quantum critical point.
Physically, the opening of the gap shifts spectral weight from low frequencies $\omega \sim \Gamma$ in the normal phase, where $\Gamma$ is
the renormalized single-particle scattering rate defined in Eq.~\ref{eq:Gamma}, to higher frequencies $\omega \sim \Delta$
in the superconducting phase. This shift of spectral weight transforms the quasi-elastic peak of the normal state into a resonance peak in
the superconducting state. Close to the nematic quantum critical point the spectral weight of this nematic resonance quickly overwhelms
that of the pair-breaking peak and grows as the critical point is approached as a function of the tuning parameter such as
doping\cite{Gallais2015}. The presence of a full gap is however crucial for the existence of the resonance, as nodes will effectively wash
out the divergence of the real part of the bare susceptibility $\Pi$. For more details we invite the reader to consult Ref.~\cite{Gallais2015}.

\subsection{Experiments: Fingerprints of a nematic resonance in superconducting Ba(Fe$_{1-x}$Co$_{x}$)$_2$As$_2$}
\label{subsec:nematic-resonance-expt}

\begin{figure}
\centering
\includegraphics[clip,width=0.99\linewidth]{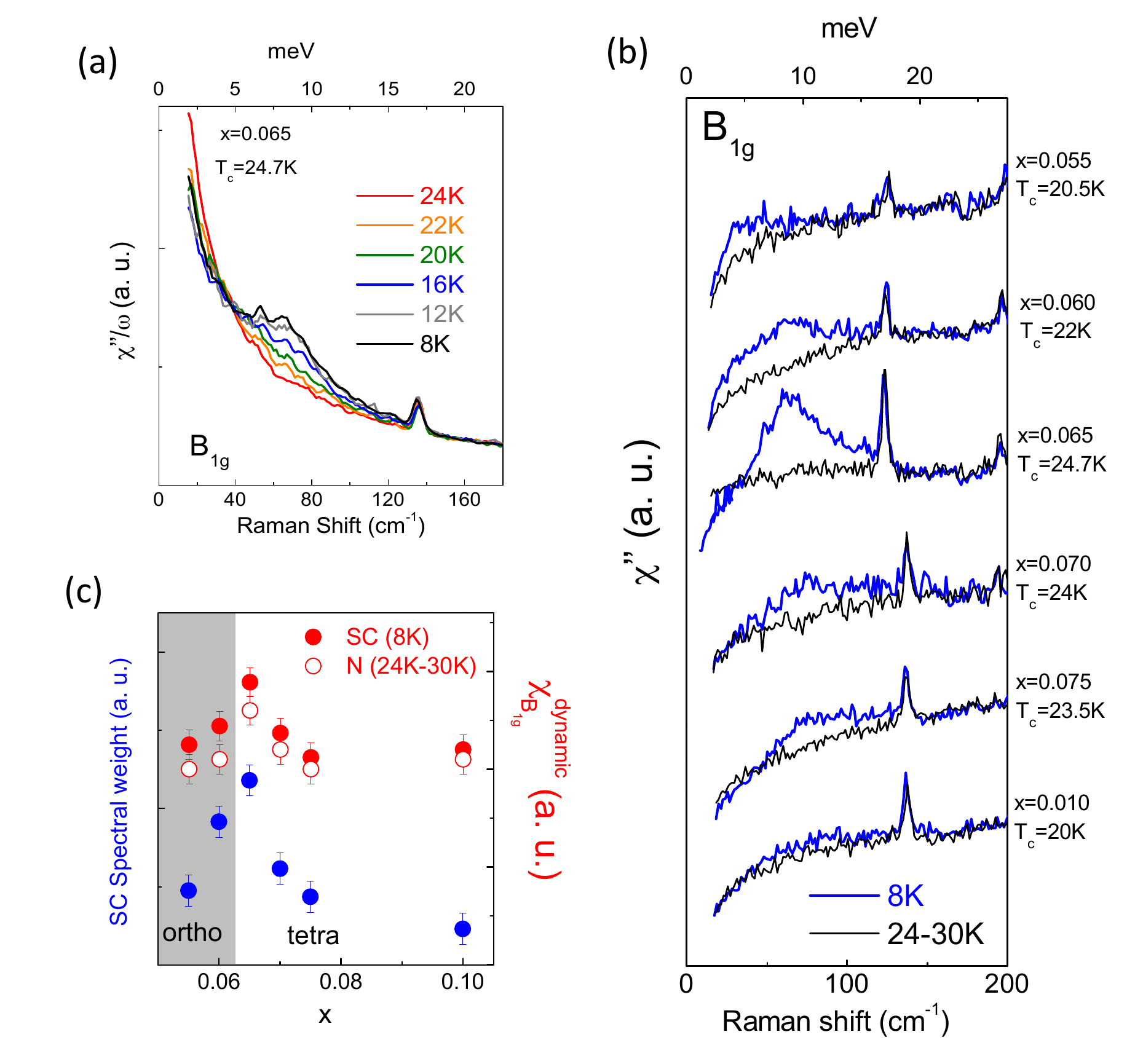}
\caption{(a) Evolution the $B_{1g}$ Raman conductivity $\chi''/\omega$ across $T_c$ for Co-Ba122 ($x$=0.065) \cite{Chauviere2010}.
(b) Co doping evolution of the SC $B_{1g}$ Raman response (in blue). The responses just above $T_c$ are shown in black. (c) Evolution of
the integrated SC spectral weight of the Raman response  $\chi''$ (in blue) \cite{Chauviere2010} as a function of Co doping. The corresponding
nematic susceptibility $\chi_{B_{1g}}^{dynamic}$ both at (N) and well below $T_c$ (SC) are also shown in red.}
\label{fig11}
\end{figure}

Raman scattering data provide a direct insight into this problem by following the evolution of the quasi-elastic peak upon crossing $T_c$.
Of particular interest are sample compositions which shows significant nematic fluctuations down to $T_c$ while staying in the tetragonal
phase. This is illustrated for optimally doped Co-Ba122 ($x$=0.065) in fig. \ref{fig11}(a). At this composition the low energy $B_{1g}$ QEP
observed in the normal state (see fig. \ref{fig78}) is strongly renormalized below $T_c$
with a suppression at low energy and the emergence of a well-defined peak at finite energy in the SC state.

The SC peak is only observed in $B_{1g}$ symmetry. It was initially interpreted as a Cooper pair-breaking peak associated to the creation
of pairs of Bogoliubov quasiparticles in a BCS
state \cite{Muschler2009,Chauviere2010,Sugai2010,Mazin2010,Setty2014}. Its energy is indeed close to twice the energy gap 2$\Delta$ detected
by ARPES experiments at the electron pockets in similarly doped
crystals \cite{Terashima2009,Hajiri2014}. Its Co doping dependence however displays a striking departure from simple BCS-like expectations
(see fig. \ref{fig11}(b)). Coming from the tetragonal overdoped side ($x$=0.1, $T_c$=20~K) the SC peak spectral weight is strongly enhanced
upon approaching the boundary between the tetragonal and orthorhombic phases which located between $x$=0.065 and $x$=0.06 in Co-Ba122. In
the nematic / SDW phase the peak spectral weight decreases and a much weaker peak emerges at a smaller energy \cite{Chauviere2010}.
\par
The strong enhancement observed on the tetragonal side ($x>$0.06) of the phase diagram is at odds with the small changes in $T_c$ when going
from $x$=0.1 ($T_c$=20~K) to $x$=0.065 ($T_c$=24.7~K). It is also inconsistent with the fact that the peak energy actually softens
towards $x$=0.065 \cite{Chauviere2010} \footnote{In  the BCS framework the spectral weight of the Raman pair-breaking should scale as $\Delta$.}.
On the other hand the doping dependence of the peak spectral weight tracks the behavior of the nematic response above $T_c$: it is stronger
for the Co compositions when the nematic susceptibility is the strongest at $T$=$T_c$, i.e. close to the nematic instability
(see Fig. \ref{fig11}(c)). This observation indicates that, as expected from the simple RPA picture discussed above, the same interaction
responsible for the enhancement of the nematic susceptibility is also enhancing the SC peak spectral weight producing a nematic resonance
close to the nematic critical point. We note that a similar enhancement of a $B_{1g}$ SC peak spectral weight has also been recently observed
in electron doped Co-Na111 upon approaching the nematic instability, indicating that the effect is possibly generic to electron doped FeAs
systems \cite{Thorsmolle2015}. This implies that the observed peak in the superconducting $B_{1g}$ channel is due to the nematic resonance
rather than simply non-interacting pair-breaking physics.

\par
While the nematic resonance is essentially a delta function for a clean system, finite lifetime effects due to e.g. disorder are expected to
broaden it. The broadening can be especially significant if the resonance is not fully detached from the $2\Delta$ pair-breaking peak. It is
interesting to note the peak observed in the experiments on Co-Ba122 are significantly broader than in Co-Na111: at least 5~meV (full width
at half maximum) for the former while it can be as sharp as 1~meV for the latter \cite{Muschler2009,Chauviere2010,Thorsmolle2015}. This
indicates that a fully detached and well-defined resonance is only present in Co-Na111, while the nematic resonance manifests itself only
as a enhanced pair-breaking peak in Co-Ba122. This may reflect a stronger coupling $g$ in the case of Co-Na111, but could also be due to a
larger disorder due to the higher Co concentration of optimally doped Co-Ba122.
\par
In principle the spectral weight of the resonance is expected to diverge and its energy to go to zero at the nematic quantum critical
point $x$=$x_c$. This corresponds to the situation where the nematic coupling reaches the critical value $g_c$ such that the Stoner
criterion is satisfied $r_0$=0 (see Eq. \ref{eq:r0}). However because the nematic Raman response does not couple to the lattice,
this does not happen: the actual thermodynamic QCP occurs before the bare nematic critical point $x_c$: $x_{QCP}>x_c$ or $g_{QCP}<g_c$.
In Co-Ba122 $x_c$ can be estimated by extrapolating the doping dependence of $T_0$: $x_c\sim$ 0.055 \cite{Gallais2013,Gallais2015}.
$x_{QCP}$ on the other hand lies at a slightly higher doping: between $x$=0.06 and $x$=0.065 (see Fig. \ref{fig5}(b)) . Thereforejust like
the divergence of the nematic susceptibility in the normal state is preempted by the structural transition which occurs at a higher
temperature $T_S>T_0$, the complete softening of the resonance energy is also preempted by the nematic order which
intervenes at a higher doping. This also explains the relatively mild softening of the mode energy seen in the experiments in Co-Ba122.

\subsection{Relationship with other superconducting collective modes}
\label{subsec:sc-collective-mode}

The nematic resonance observed in the Raman spectrum of electron doped Fe SC bears a strong analogy with neutron resonance observed in
several families of unconventional SC. In both cases they are due to residual electronic interactions in the particle-hole channel, which
within RPA type pictures, trigger a well-defined excitonic-like pole in the charge and spin SC responses respectively. The nature of the
residual interactions differs: magnetic and centered at finite $q_{AF}$ for the neutron resonance while they are centered at zero wavevector
for the nematic resonance. Both resonances are connected to soft excitations associated to nearby quantum critical points: antiferromagnetic
in one case and nematic in the other. It so happens that both critical points seem to exist in Fe SC and therefore both neutron and nematic
resonance are present in their SC state.
There is another key difference between both resonances: while the neutron resonance requires a sign-changing gap, the nematic resonance
only requires the presence of a full gap. Because the presence of nodes is suspected in several Fe SC systems, it is likely that the nematic
resonance will not be observed in several systems. The presence or not of a nematic resonance near the nematic instability is thus a powerful
test for the presence of nodes for several systems including e.g. P-doped Ba122.
\par
Finally we should note that other collective modes have been predicted to occur in the SC state of Fe
SC \cite{Chubukov2009b,Lee-Zhang2009,Scalapino2009,Ong2013,Marciani2013}. Among them the Bardasis Schrieffer (BS)
mode \cite{Bardasis,Lee-Zhang2009,Scalapino2009,Maiti2015}, an in-gap electron bound state in a sub-leading pairing channel, is the most
relevant for our discussion because of the near degeneracy of $s$ and $d$-wave paring state in many Fe
SC \cite{Kuroki2008,Graser2009,Wang2009,Graser2010,Thomale2011,Maiti2011b, Maiti2011}. Fingerprints of its presence have been reported in
the Raman spectra of optimally hole doped K-Ba122 where several sharp modes are detected below $T_c$ \cite{Kretzschmar2013,Bohm2014}. In
principle both the nematic resonance and the BS mode can be simultaneously present in the SC Raman spectrum and may even
couple \cite{Lee2013,Khodas2014}. However it is likely that their respective visibility will be strongly system dependent.
While anisotropic gaps will suppress both collective modes efficiently, their presence is also linked to different
criteria: the nematic resonance relies on the proximity of a nematic quantum critical point while the BS mode will occur when $s$ and $d$-wave
pairing states are nearly degenerate \cite{Monien90,Scalapino2009}. In this respect we note that, from the point of view of nematicity,
shear modulus and Raman scattering data shows much weaker nematic fluctuations near optimally hole doped K-Ba122 where a region of $C_4$
magnetic phase has also been observed \cite{Bohmer2014,Massat2015}. We therefore suspect that,
by contrast to the electron doped side, the nematic resonance may not be present in hole-doped 122 systems at least near optimal doping.
This illustrates the remarkable richness of Fe SC where the nature of the  SC ground state and its collective excitation spectrum can
vary strongly from one system to the other.

\section{Conclusions}

The purpose of this review was two-fold. First, we have provided a general theoretical framework within which one can understand
electronic Raman scattering measurements near a nematic instability. Since Raman scattering measurement in the appropriate symmetry
couples directly to the nematic order parameter involving the charge degrees of freedom, we have argued that it is one of the
few direct probes of nematic fluctuations in a metallic system. However, contrary to thermodynamic probes like elastic constant
measurements, the electronic
Raman response probes the frequency-momentum dependent nematic susceptibility in its dynamical limit. An important consequence
of this is that it is essentially blind to the lattice strains and the acoustic phonons. This, in turn, implies that Raman response
is an unique probe of pure electronic nematic correlations. Experimental data on electron doped 122, and more
recently 111, Fe SC systems are consistent with this theoretical picture, and therefore provide unambiguous evidence of an enhanced
nematic susceptibility of electronic origin. The comparison with shear modulus data confirm the absence of coupling between the electronic
Raman response and the lattice strains. It also shows that the nematic fluctuations observed in the charge sector by Raman scattering can
account for the observed lattice softening in electron doped 122, providing further evidence that the structural
phase transition is electronic driven.
\par
However, despite the encouraging consistency between different probes of nematicity in the tetragonal phase,
there remains several unsolved questions. The first one is related to the
issue whether the electronic nematicity is driven by spin, orbital or charge degrees of freedom.
In this respect FeSe is an interesting system because
of the absence of magnetic ordering. While preliminary Raman measurements indicate the presence of charge nematic fluctuations above
$T_S$ in FeSe \cite{GallaisMarchMeeting,Massat2015b}, the nature of the magnetic fluctuations in this system remains to
be clarified  \cite{Bohmer2015, Rahn2015,Wang2015}. The second question is how generic are nematic fluctuations in Fe SC and
whether or not
they play a role in the SC pairing mechanism. Recent transport measurements advocate for the presence of underlying nematic quantum
critical point near optimal doping for several systems and dopings. This will have to be confirmed by other probes and we believe
electronic Raman scattering is ideally suited to answer this question. Indeed we have shown that the presence of a nematic resonance
in the SC Raman response is a natural consequence of the proximity to a nematic quantum critical point, providing a smoking-gun
experiment for its existence.
\par
From a more general perspective, the discovery of nematicity in Fe SC, first revealed via transport measurements, has stimulated
intense efforts to design and understand experiments capable of probing nematic degrees of freedom. These efforts can now be
capitalized to search for other systems where nematic degrees of freedom have been either predicted or indirectly observed. We have
already mentioned cuprates and bi-layer ruthenates in the introduction, we can also add less correlated systems like bi-layer
graphene \cite{Lemonik2010,Vafek2010,Mayorov2011} and even bismuth \cite{Kuchler2014,Collaudin2015}. The presence of nematic correlations
can also be used to tune electronic orders via uniaxial stress or strain, just like magnetic field is routinely used to tune
magnetic orders. Strain control of electronic phases has already been demonstrated in semiconductor based heterostructures like
AlAs \cite{Shkolnikov2005, Gunawan2006}. In more correlated materials strain-effects have been demonstrated recently in
Sr$_2$RuO$_4$ where a strong $T_c$ enhancement under uni-axial strain was observed \cite{Hicks2014}. Up to now the microscopic origin
of these kind of
effects remain relatively poorly studied however. It is therefore likely that nematicity will remain a subject of intense research
in the coming years.

\section*{Acknowledgements}
For the experimental results presented in this review we are indebted to the works of three PhD students, namely
L. Chauvi\`ere, Y. -X. Yang
and P. Massat. We acknowledge D. Colson and A. Forget whose high-quality single crystals have been instrumental to this work, and
we acknowledge F. Rullier-Albenque for sharing with us transport data. We also acknowledge R. M. Fernandes and J. Schmalian for very fruitful collaborations. Y. G. would like to thank his close collaborators within the Raman group at Universit\'e Paris Diderot for their constant feedback and various technical help, namely M. A. M\'easson, M. Cazayous and
A. Sacuto. We are also grateful to E. Bascones, L. Benfatto, G. Blumberg, A. B\"{o}hmer, V. Brouet, A. Cano, R. Hackl, P. Hirschfeld, K. Ishida,
M. H. Julien, H. Kontani, D. Maslov, P. Toulemonde, and B. Valenzuela for many helpful conversations. Part of this work was funded by
Agence Nationale de la Recherche through ANR Grant PNICTIDES and by a SESAME Grant for R\'egion Ile-de-France.


\bibliographystyle{unsrt}
\bibliography{bibliographie}


\end{document}